\documentclass[aps,prd]{revtex4}
\usepackage{epsfig,epsf}
\usepackage{amsmath}
\usepackage{amsthm}
\usepackage{amsfonts}
\usepackage{amssymb}
\usepackage{dsfont}
\usepackage{multirow}
\usepackage{appendix}
\usepackage{slashed}
\usepackage[active]{srcltx}
\usepackage{psfrag}
\usepackage{subfigure}
\usepackage{booktabs}
\usepackage[switch,modulo]{lineno}
\usepackage[colorlinks,citecolor=blue,urlcolor=red,linkcolor=red]{hyperref}
\usepackage[colorlinks,citecolor=blue,urlcolor=red,linkcolor=red]{hyperref}
\usepackage{graphicx}
\usepackage{epstopdf}
\usepackage{rotating}
\usepackage{booktabs}
\usepackage{wasysym}

\begin{document}

\title{{\Large{\bf  Semileptonic Decays of Heavy Vector Mesons in Hard-Wall AdS/QCD with $N_{f}=5$}}}
\author{
\small S. Momeni $^1$  \footnote {e-mail: samira.momeni@ph.iut.ac.ir },
\small M. Saghebfar $^2$ \footnote {e-mail: saghebfar@mut-es.ac.ir},
\small A. Ranjbar $^3$  \footnote {e-mail: Aminranjbar.sub@gmail.com }}

\affiliation{\emph{ $^1$ Department of Physics,Malek Ashtar
University of  Technology, Shahin Shahr , Iran \\
$^2$Optics-Laser Science and Technology Research Center, Malek Ashtar
University of  Technology, Shahin Shahr , Iran\\
$^3$ Department of Physics, Shahid Beheshti University, Tehran, Iran \\
}}
\begin{abstract}
We present a systematic study of semileptonic decays of heavy vector mesons within the Hard-Wall AdS/QCD framework. Semileptonic decays provide a clean environment for testing the Standard Model because the leptonic current is well understood, while the hadronic part is encoded in transition form factors that reflect non-perturbative QCD dynamics. The processes under consideration include:
\( \psi \to D^{*-} ( D^{-} ) \ell^{+} {\nu}_{\ell} \), \( D^{*0} \to \pi^{-} \ell^{+} {\nu}_{\ell} \), \( D^{*+}_{s} \to K^{0} \ell^{+} {\nu}_{\ell} \), \( D^{*0} \to \rho^{-} \ell^{+} {\nu}_{\ell} \), \( B_c^{*+} \to B^{*0} \ell^{+} {\nu}_{\ell} \),  \( B_c^{*+} \to \psi \,\ell^{+} {\nu}_{\ell} \), \( B^{*-} \to D^{*0} \ell^{-} \bar{\nu}_{\ell} \), \( B^{*0} \to D^{*+}(D^{+}) \ell^{-} \bar{\nu}_{\ell} \), \( B^{*0}_{s} \to D^{*+}_{s} (D^{+}_{s}) \ell^{-} \bar{\nu}_{\ell} \), \( \psi \to D^{*-}_{s}(D^{-}_{s}) \ell^{+} {\nu}_{\ell} \), \( D^{*0} \to K^{*-} (K^{-}) \ell^{+} {\nu}_{\ell} \), \( B_c^{*-} \to D^{*0} \ell^{-} \bar{\nu}_{\ell} \), \( \bar{B_{s}}^{*0} \to K^{*+} \ell^{-} \bar{\nu}_{\ell} \), and \( \bar{B}^{*0} \to \rho^{+} \ell^{-} \bar{\nu}_{\ell} \).
These decays correspond to four different quark-level transitions: \(c\to d\), \(b\to c\), \(c\to s\), and \(b\to u\), allowing us to explore a broad range of flavor dynamics.\\
Using the Hard-Wall AdS/QCD model, we compute the relevant transition form factors as functions of the momentum transfer \(q^{2}\). The framework provides analytic expressions for meson wave functions and decay constants, which serve as inputs for the form-factor calculations. From these form factors, we evaluate the differential and integrated branching ratios for each decay mode. Where possible, we compare our predictions with results obtained from other non-perturbative approaches. 
\end{abstract}


\maketitle

\section{Introduction}\label{sec.1}
The study of decays of pseudoscalar bottom ($B$) and charmed ($D$) mesons is essential for advancing fundamental physics. These decays serve as precise probes of the Standard Model (SM), offering insights into possible new physics and addressing key cosmological questions. Processes such as \(B^0 \to D^{(*)-}\ell^+\nu_\ell\), \(B_s \to D_s^{(*)}\ell\nu_\ell\), and \(D^0 \to K^-\ell^+\nu_\ell\) provide a laboratory for measuring Cabibbo-Kobayashi-Maskawa (CKM) matrix elements (e.g., \(|V_{cb}|\) and \(|V_{ub}|\)), studying charge-parity (CP) violation within the SM, and describing neutral \(D\)-\(\bar D\) and \(B\)-\(\bar B\) mixing. Accurate extraction of these parameters tests the unitarity of the CKM matrix-a cornerstone of the SM-and any deviation could signal physics beyond the current framework. Equally important are tests of lepton flavor universality (LFU).\\
In addition to pseudoscalar $D$ mesons, the weak decays of vector \(D^{*}\) mesons offer another valuable platform for investigating the properties of these states, exploring their decay mechanisms, and testing the Standard Model (SM). However, owing to the small strength of the weak interaction, such decays are typically very rare processes \cite{CLFQM2024}.

A similar situation holds for charmonium states like the $\psi$  particle. While its strong and electromagnetic decays have been extensively studied both experimentally and theoretically for decades, weak decays of the $\psi$ remain largely unexplored. The first estimates placed the branching ratios of these weak processes at the order of \(10^{-8}\) \cite{Sanchis1994}. At that time, such tiny rates drew little attention, as the available data samples were far from achieving the required sensitivity. Consequently, little further research was conducted on this topic for a long period.

Due to recent advances in accelerator and detector technologies, more precise measurements have become feasible, reigniting interest in the weak decays of the $\psi$. The BES collaboration has initiated measurements of some rare weak decays and has set an upper bound on the branching ratio of \(\psi \to D^+ e^+ \nu_e\) at the order of \(10^{-5}\), based on the existing $\psi$ data sample \cite{Ablikim2006}. The forthcoming upgraded BESIII detector is expected to accumulate approximately \(10^7\) $\psi$ events per year, corresponding to \(5.8 \times 10^{10}\) \(\psi\) decays annually \cite{Harris20093}. This would make it marginally possible to observe such weak decays-or at least to expect non-zero signal events. Hence, more careful theoretical investigations of these decays are now warranted.

In contrast to their pseudoscalar counterparts, vector bottom mesons ($B^{*}$) remain experimentally underexplored (see Ref. \cite{pdg}), largely due to their lower production rates and detection efficiencies. The dominant decay mode of the $B^{*}$ meson is the electromagnetic process $B^{*} \to B \gamma$. Nevertheless, rapid advances in experimental particle physics in recent years offer hope for significant improvement, particularly with future data from the LHC and Belle-II \cite{pdg, Chang220005, Kumar2018}. For instance, Belle-II is expected to reach an annual integrated luminosity of approximately $13\,\text{ab}^{-1}$, which could enable the observation of $B^{*}$ weak decays with branching fractions above $\mathcal{O}(10^{-9})$ \cite{Chang220005, Sahoo2017}. Furthermore, the LHC experiments will provide extensive information on $B^{*}$ weak decays, thanks to the substantially larger beauty production cross section in $pp$ collisions compared to $e^{+}e^{-}$ collisions \cite{Xu2016}. Therefore, timely theoretical studies of $B^{*}$ weak decays are urgently needed to provide useful predictions and guidance for forthcoming measurements.

In this paper, we focus on the semileptonic decays of $B^{*}$, $D^{*}$, and $\psi$ into light and heavy vector or pseudoscalar mesons.
These decays are investigated using a variety of theoretical approaches, including the three-point QCD sum rule (3PSR)  \cite{Wang3PSR}, the covariant constituent quark model (CCQM) \cite{Ivanov2015CCQM}, the covariant light-front quark model (CLFQM) \cite{CLFQM20083, CLFQM2024, CLFQM20242, FQCD2019, CLFQMVV}, the light-cone QCD sum rule (LCSR) \cite{Li2020}, lattice QCD (LQCD) \cite{LQCD2008}, naive factorization (NF) \cite{Nfact}, the light-front quark model (LFQM) \cite{LFQM2019}, and the Standard Model (SM) \cite{BtoD2016}.

In this paper, we employ the Hard-Wall anti-de Sitter/quantum chromodynamics (AdS/QCD) framework  to study $V \to V(P)\ell^{+}\nu$ decay. 
 This approach connects the non-perturbative regime of four-dimensional QCD to a weakly coupled five-dimensional gravitational theory in AdS space \cite{Erlich2005, Rold2005}. The AdS/QCD duality, rooted in the gauge/gravity correspondence, allows researchers to apply tools from string theory and general relativity to challenging strong-coupling problems in QCD \cite{Maldacena1998}. In the Hard-Wall model, the AdS space is compactified by imposing boundary conditions on the radial coordinate \(z\), with \(\varepsilon (\to 0) \leq z \leq z_{0}\). The lower limit corresponds to the ultraviolet (UV) boundary of QCD, while the infrared cutoff \(z_{0}\) captures confinement effects.
 
The framework has proven highly versatile in phenomenological studies, enabling the extraction of meson and baryon masses, decay constants, and parton distribution functions \cite{Sonnenschein2007}. It also naturally accommodates multiple quark flavors, facilitating investigations of flavor mixing, heavy quark dynamics, and other complex phenomena \cite{Karch2006b}. By providing access to strongly coupled regimes where perturbative methods fail, AdS/QCD offers valuable insights into hadron dynamics and the strong force \cite{Brodsky2006}. Moreover, AdS/QCD predictions can be directly compared with experimental data, making the framework a useful guide for experimental searches and a tool for refining theoretical models \cite{Gursoy2008}. Its adaptability extends to a wide range of processes, including semileptonic decays, electromagnetic interactions, and CP violation studies, highlighting its broad applicability in particle physics \cite{Pomarol2008}. With \(N_f = 3\), this approach has been used to estimate numerous phenomenological quantities, such as masses, decay constants, electromagnetic and gravitational form factors, and distribution amplitudes (DAs) for light vector, axial-vector, and pseudoscalar mesons. In addition, parton distribution functions and transverse momentum distributions (TMDs) have also been analyzed \cite{Grigoryan2007, Grigoryan20072, Grigoryan2008, Kwee2008, Kwee20082, Boschi2006, Abidin2008, Abidin20082, Abidin20083, Abidin20091, Abidin2009, Forshaw2011, Ahma1, Ahma2, Ahma3, Ahma4, Ahma5, AhmaLord, ChaBro, Momeni2017, Momeni2018, Ahmady2019, Ahmady2020}.

Extending the Hard-Wall holographic QCD framework to \(N_f = 4\), researchers have studied the strong couplings of various vertices, including \((\rho^{n}, \rho, \rho)\), \((\rho^{n}, K, K)\), \((\rho^{n}, K^{*}, K^{*})\), \((\rho^{n}, D, D)\), \((\rho^{n}, D^{*}, D^{*})\), \((D^{(*)}, D, A)\), \((D^{(*)}, D^{(*)}, V)\), \((D_{1}, D_{1}, P)\), \((\psi, D, D^{(*)}, P)\), and \((\psi, D, D^{(*)}, A)\) \cite{Bayona2017, Momeni2021}. Furthermore, transition form factors for semileptonic decays \(D \to (V, A, S)\,\ell\,\nu_{\ell}\) have also been calculated \cite{Momeni2022}. In Ref. \cite{Momeni2023}, heavy meson couplings were analyzed within the Hard-Wall AdS/QCD model with \(N_f = 5\).
In this paper, we compute the transition form factors for the semileptonic decays of heavy vector mesons into vector or pseudoscalar mesons using the Hard-Wall AdS/QCD framework.
This paper is organized in the following manner. Section \ref{sec.2} describes the Hard-Wall AdS/QCD framework, including the treatment of scalar, pseudoscalar, vector, and axial-vector mesons. Within this section, we derive the wave functions and decay constants for these mesons, as well as 
the transition form factors  for the semileptonic decays under investigation. Section \ref{sec.3} is dedicated to the numerical analysis of these form factors and the corresponding branching ratios. To assess the reliability of our predictions, we compare them with results obtained from other theoretical methods.  Finally, Section \ref{sec.5} presents our conclusions and offers further discussion.

\section{ $V \to V(P)$ transitions  in the Hard-Wall AdS/QCD Model }\label{sec.2}

In this section, we present the theoretical framework for calculating the form factors of the \( V \to V\) and \( V \to P \) transitions  within the Hard-Wall AdS/QCD model. We first introduce the representation of mesons in this framework, establishing the necessary groundwork for the analysis. Subsequently, we derive the form factors for the specified decays, explicitly outlining the computational methodology and underlying assumptions.  

\subsection{Meson Structure and Dynamics in the Hard-Wall AdS/QCD Model}
This section presents a systematic investigation of both heavy and light mesons within the Hard-Wall AdS/QCD framework. We begin by establishing the fundamental geometric structure through the five-dimensional AdS space metric:

\begin{equation}\label{eq.1}  
ds^2 = \frac{R^2}{z^2} \left( \eta_{\mu\nu} dx^\mu dx^\nu - dz^2 \right), \quad (\varepsilon \leq z \leq z_0).
\end{equation}
where the four-dimensional Minkowski metric \(\eta_{\mu\nu} = \mathrm{diag}(1, -1, -1, -1)\) and the AdS curvature radius \(R\) (normalized to unity for pure AdS space) form the basis of our geometric description.

The AdS/QCD correspondence dictates a direct relationship between bulk fields in \(\mathrm{AdS}_5\) and QCD operators. Our framework incorporates a scalar field \(X\) dual to the quark condensate operator \(\bar{q}_L q_R\), along with gauge fields \(L^{\mu,a}\) and \(R^{\mu,a}\) corresponding to the chiral currents \(J^{\mu,a}_{L/R} = \bar{q}_{L/R} \gamma^\mu t^a q_{L/R}\). The chiral quark fields \(q_{L/R} = (1 \pm \gamma_5)q\) and \(SU(N_f)\) generators \(t^a\) (with \(N_f = 5\) to encompass all relevant quark flavors) complete the field content.
The dynamics are governed by the \(SU(5)_L \otimes SU(5)_R\)-invariant action:
\begin{equation}\label{eq.2s}  
S = \int d^5 x \, \sqrt{g} \, \mathrm{Tr} \left\{ |D_M X|^2 + 3|X|^2 - \frac{1}{4g_5^2} \left( L^{MN} L_{MN} + R^{MN} R_{MN} \right) \right\},
\end{equation}

where the covariant derivatives and field strength tensors exhibit the standard non-Abelian structure. The decomposition into vector (\(V = (L + R)/2\)) and axial-vector (\(A = (L - R)/2\)) components reveals the chiral symmetry breaking pattern.
The scalar sector features a background solution \(X_0(z)\) with explicit dependence on quark masses and condensates:
\begin{equation}\label{eq.4x0}  
2 X_{0,ij}(z) = \zeta M_{ij} z + \frac{\Sigma_{ij}}{\zeta} z^3,
\end{equation}

where the matrices \(M = \mathrm{diag}(m_u, m_d, m_s, m_c, m_b)\) and \(\Sigma = \mathrm{diag}(\sigma_u, \sigma_d, \sigma_s, \sigma_c, \sigma_b)\) encode the flavor structure, while \(\zeta = \sqrt{N_c}/2\pi\) provides proper normalization.
The quadratic expansion of the action yields the mesonic wavefunctions and mass spectra:
\begin{eqnarray} \label{eq.5}
S=\int\, d^5 x  \bigg\{\sum_{a=1}^{15} \frac{-1}{4g_5^2 z} \eta^{MM'}\,\eta^{NN'}\,(\partial_M V^a_N-\partial_N V^a_M)\,
(\partial_{M'} V_{N'a}-\partial_{N'} V_{M' a})+\frac{ {M_V^a}^2}{2z^3}\eta^{MM'}{V_{M a}}{V^a_{M'}}\nonumber\\
-\frac{1}{4g_5^2 z}  \eta^{MM'}\,\eta^{NN'}(\partial_M A^a_N-\partial_N A^a_M)\,(\partial_{M'} A_{N' a}-\partial_{N'}
A_{M' a})+\frac{{M_A^a}^2}{2z^3}\eta^{MM'}(\partial_M\pi^a-A^a_M)\,(\partial_{M'}\pi_{b}-A_{M' b}) \bigg\},\nonumber
\end{eqnarray}

with the vector and axial-vector masses determined through:
\begin{align}  
M_V^{a2} &= -2 \, \mathrm{Tr} \left( [t^a, X_0]^2 \right),\label{eq.defmva} \\  
M_A^{a2} &= 2 \, \mathrm{Tr} \left( \{t^a, X_0\}^2 \right).\label{eq.defmvaa}
\end{align}

The mass spectrum of vector and axial-vector mesons in our framework is characterized by the squared masses \( {M_V^a}^2 \) and \( {M_A^a}^2 \) for \( a=1,\cdots, 24 \), as calculated in \cite{Momeni2023}. The complete methodology for deriving meson wavefunctions and decay constants - encompassing vector, axial-vector, pseudoscalar, and scalar states - from the action in Eq. (\ref{eq.5}) has been established in previous works \cite{Abidin2009,Abidin2008,Abidin20082,Momeni2021,Momeni2022}.
The dynamical equations for the vector (\( V^a_M \)) and axial-vector (\( A^a_M \)) fields follow from variational principles:
\begin{equation}
\eta^{ML}\partial_M\left(\frac{1}{z}\left(\partial_L V^a_N - \partial_N V^a_L\right)\right) + \frac{\alpha^a(z)}{z}V^a_N = 0,
\label{eq.eomva1}
\end{equation}
\begin{equation}
\eta^{ML}\partial_M\left(\frac{1}{z}\left(\partial_L A^a_N - \partial_N A^a_L\right)\right) + \frac{\beta^a(z)}{z}A^a_N = 0.
\label{eq.eomva2}
\end{equation}

The coefficients in the equations of motion are given by:
\begin{equation}
\alpha^a(z) = \frac{g_5^2 {M_V^a}^2}{z^2}, \quad \beta^a(z) = \frac{g_5^2 {M_A^a}^2}{z^2}, 
\end{equation}

where the vector and axial-vector meson masses are given by the constant eigenvalues ${M_V^a}^2$ and ${M_A^a}^2$ (from Eqs. \ref{eq.defmva} and \ref{eq.defmvaa}), whose $z$-dependence is dictated by the AdS metric. The bulk mass contributions for the $a^{th}$ flavor are encoded in the holographic profiles $\alpha^a(z)$ and $\beta^a(z)$. The five-dimensional coupling constant is fixed by the QCD color number as $g_5^2 = {12\pi^2}/{N_c}$.\\
The vector and axial-vector fields decompose as:
\begin{eqnarray}
V^a_N =(V^a_\mu, V^a_z),~~~ V^a_\mu = V^a_{\mu\perp} + V^a_{\mu\parallel}, \\
A^a_N = (A^a_\mu, A^a_z),~~~A^a_\mu = A^a_{\mu\perp} + A^a_{\mu\parallel},
\end{eqnarray}

with the axial gauge condition $A^a_z = 0$ eliminating the redundant degree of freedom while maintaining gauge invariance.
The transverse components \( V^a_{\mu\perp} \) and \( A^a_{\mu\perp} \) satisfy  equations of motion in the holographic coordinate \( z \) as:
\begin{eqnarray}
\left(\partial_z\frac{1}{z}\partial_z+\frac{q^2-\alpha^a}{z}\right) V^a_{\mu\perp}(q,z)=0,\label{eq.eomv2}\\
\left(\partial_z\frac{1}{z} \partial_z+\frac{q^2-\beta^a}{z}\right) A^a_{\mu\perp}(q,z)=0,\label{eq.eoma2}
\end{eqnarray}

where \( q^\mu \) represents the four-dimensional momentum conjugate to the space-time coordinates \( x^\mu \). The transversality conditions \( \partial^\mu V^a_{\mu\perp} = 0 \) and \( \partial^\mu A^a_{\mu\perp} = 0 \) are essential in deriving Eqs. (\ref{eq.eomv2}) and (\ref{eq.eoma2}).
The transverse components of the vector and axial-vector fields, \( U = (V, A) \), can be expressed as
\begin{eqnarray}
U^a_{\mu\perp}(q,z)= U^{0a}_{\mu\perp}(q) {\cal U}^a(q^2,z),
\end{eqnarray}

where \( U^{0a}_{\mu\perp}(q) \) is the boundary value at the UV and \( \mathcal{U}^a(q^2, z) \) is the bulk-to-boundary propagator, which satisfies the same equations of motion as \( U^a_{\mu\perp}(q, z) \) with the boundary conditions \( \mathcal{U}^a(q^2, \varepsilon) = 1 \) (UV) and \( \partial_z \mathcal{U}^a(q^2, z_0) = 0 \) (IR).

We now consider the longitudinal components of the fields. For the vector sector, the scalar modes are parametrized as \( V^a_{\mu\parallel} = \partial_\mu \xi^a \) and \( V^a_z = -\partial_z \tilde{\pi}^a \), which satisfy the coupled system:
\begin{eqnarray}
-q^2\partial_z \tilde \phi^a(q^2,z) +\alpha^a\partial_z \tilde \pi^a(q^2,z)=0\,,\label{eq.eomv3}\\
\partial_z \bigg(\frac{1}{z}\partial_z\tilde \phi^a(q^2,z) \bigg)
 -\frac{\alpha^a}{z}\big(\tilde \phi^a(q^2,z)-\tilde \pi^a(q^2,z)\big)=0,\label{eq.eomv4}
\end{eqnarray}

where \(\xi^a = \tilde{\phi}^a - \tilde{\pi}^a\). The boundary conditions are fixed as \(\tilde{\phi}^a(\varepsilon) = 0\) and \(\tilde{\pi}^a(\varepsilon) = -1\) at the UV boundary \(z=\varepsilon\), with Neumann conditions \(\partial_z \tilde{\phi}^a(z_0) = \partial_z \tilde{\pi}^a(z_0) = 0\) at the IR boundary \(z=z_0\).
For the pseudoscalar sector, the fields \(\pi^a\) and \(\phi^a\) (where \(A^{a}_{\mu\parallel} = \partial_\mu \phi^a\)) obey:
\begin{eqnarray}
-q^2\partial_z \phi^a(q^2,z)+\beta^a(z)\partial_z \pi^a(q^2,z)=0\,,\label{longtueq1}\\			
\partial_z \left(\frac{1}{z}\partial_z\phi^a(q^2,z) \right)  -
\frac{\beta^a(z)}{z}\left(\phi^a(q^2,z)-\pi^a(q^2,z)\right)=0\,.\label{longtueq2} 					
\end{eqnarray}

with analogous boundary conditions: \(\phi^a(\varepsilon) = 0\), \(\pi^a(\varepsilon) = -1\) (UV), and \(\partial_z \phi^a(z_0) = \partial_z \pi^a(z_0) = 0\) (IR).

Using the Green's function formalism to solve Eqs. (\ref{eq.eomv2}--\ref{longtueq2}), the quantities \( \mathcal{V}^a \), \( \mathcal{A}^a \), \( \pi^a (\phi^a) \), and \( \tilde{\pi}^a (\tilde{\phi}^a) \) can be expressed as sums over poles corresponding to vector (\( V \)), axial-vector (\( A \)), pseudoscalar (\( P \)), and scalar (\( S \)) mesons:
\begin{eqnarray}\label{eq.aandpide}
{\cal V}^a(q^2,z)&=&\sum_n \frac{-g_5 f_{_{V^{n}}}^{a} \psi^a_{_{V^{n}}}(z)}{q^2-m_{_{V^{n}}}^2}\,,~~~
{\cal A}^a(q^2, z)=\sum_n \frac{-g_5 f_{_{A^{n}}}^{a} \psi^a_{_{A^{n}}}(z)}{q^2-{m^{a2}_{_{A^{n}}}}}\,,\label{eq.gra} \\
\phi^a(q^2, z)&=&\sum_n\frac{- g_5 {m^{a2}_{_{P^{n}}}} f_{_{P^{n}}}^a\phi^a_n(z)}{q^2-{m^{a2}_{_{P^{n}}}}}\,,~~~
\pi^a(q^2, z)=\sum_n\frac{- g_5 {m^{a2}_{_{P^{n}}}} f_{_{P^{n}}}^a\pi^a_n(z)}{q^2-{m^{a2}_{_{P^{n}}}}}\,,\label{eq.grpi}\\
\tilde \phi^a(q^2, z)&=&\sum_n\frac{- g_5 m^{a2}_{_{S^{n}}} f_{_{S^{n}}}^a \tilde \phi^a_n(z)}{q^2-{m^{a2}_{_{S^{n}}}}}\,,~~~
\tilde \pi^a(q^2, z)=\sum_n\frac{- g_5 m^{a2}_{_{S^{n}}} f_{_{S^{n}}}^a\tilde \pi^a_n(z)}{q^2-{m^{a2}_{_{S^{n}}}}}.
\label{eq.grpit}	
\end{eqnarray}

Here, $f_{_{V(A)^{n}}}^{\frac{1}{2}}$ and $f_{_{P(S)^{n}}}$ are the decay constants of the \( n^{\text{th}} \) mode of vector (axial-vector) and pseudoscalar (scalar) mesons, respectively. These decay constants are derived as follows \cite{Erlich2005}:
\begin{equation}
f^{a}_{\mathcal{Q}^n} = \gamma \left. \frac{\partial_z B^{a}_{_{\mathcal{Q}^n}}}{g_5 z} \right|_{z = \epsilon}, \label{eq.dev}
\end{equation}

where \( (B, \gamma) = (\psi, 1) \) for \( \mathcal{Q} = A, V \), and \( (B, \gamma) = (\phi \, (\tilde{\phi}), -1) \) for \( \mathcal{Q} = P \, (S) \).\\
For \( N_f = 5 \), the  vector (\( V \)), and pseudoscalar (\( \pi \)) fields-incorporating the light (\( u,d \)), strange (\( s \)), charmed (\( c \)), and bottom (\( b \)) quark states-can be decomposed into their charged components as:  
\begin{eqnarray*}
 V &=& \frac{1}{\sqrt 2}
 \left ( \begin{matrix}
  \frac{\rho^0}{\sqrt{2}} + \frac{\omega'}{\sqrt{6}} -\frac{2}{\sqrt{30}}\Upsilon  &  \rho^{_{+}}  &  K^{_{*+}}  &  \bar D^{_{*0}}& B^{_{*+}} \\
  \rho^{_{-}}   & -\frac{\rho^0}{\sqrt{2}}  + \frac{\omega'}{\sqrt{6}}  -\frac{2}{\sqrt{30}}\Upsilon   &  K^{_{*0}}  &  D^{_{*-}}& B^{_{*0}} \\
   K^{_{*-}}  &  \bar K^{_{*0}}  & - \sqrt{\frac23} \omega' -\frac{2}{\sqrt{30}}\Upsilon   &  D_s^{_{*-}}& B_s^{_{*0}} \\
   D^{_{*0}} &  D^{_{*+}}  &  D_s^{_{*+}}  &\frac{\psi}{\sqrt{2}}+\frac{3}{\sqrt{30}}\Upsilon  & B_c^{_{*+}}\\
   B^{_{*+}}&\bar B^{_{*0}} & \bar B_s^{_{*0}}& B_c^{_{*-}} &-\frac{\psi}{\sqrt{2}}+\frac{3}{\sqrt{30}}\Upsilon \\
 \end{matrix} \right ),\\
 \end{eqnarray*}
\begin{eqnarray*}
\pi &=&  \frac{1}{\sqrt 2}
\left (\begin{matrix}
  \frac{\pi^{_{0}}}{\sqrt{2}}  + \frac{\eta}{\sqrt{6}}  - \frac{2}{\sqrt{30}}\eta_b  &  \pi^{_{+}}  &  K^{_{+}}
   &  \bar D^{_{0}}& B^{_{+}} \\
  \pi^{_{-}}   & -\frac{\pi^0}{\sqrt{2}}  + \frac{\eta}{\sqrt{6}}  - \frac{2}{\sqrt{30}}\eta_b   &  K^{_{0}}  &  D^{_{-}}& B^{_{0}}\\
   K^{_{-}}  &  \bar K^{_{0}}  & - \sqrt{\frac23} \eta - \frac{2}{\sqrt{30}}\eta_b   &  D_s^{_{-}} &B_{s}^{_{0}} \\
   D^{_{0}} &  D^{_{+}}  &  D_s^{_{+}}  & \frac{\eta_c}{\sqrt{2}}+ \frac{3}{\sqrt{30}} \eta_b&B_{c}^{_{+}}\\
   B^{_{-}}  & \bar B^{_{0}} &\bar B^{_{0}}&B_{c}^{_{-}}&-\frac{\eta_c}{\sqrt{2}}+\frac{3}{\sqrt{30}}\eta_b\\
 \end{matrix} \right ),\\
 \end{eqnarray*}
\subsection{Hard-Wall AdS/QCD Calculation of $V\to V$ and $V\to P$  Form Factors}
In this section, we derive the form factors governing the semileptonic decays \( V \to V  \) and \( V \to P \) within the Hard-Wall AdS/QCD framework. These decays originate from four distinct quark-level transitions. The first, the \( c \to d \ell^{+} {\nu}_{\ell} \) channel, is exemplified by processes such as \( \psi \to D^{*-} ( D^{-} ) \ell^{+} {\nu}_{\ell} \), \( D^{*0} \to \pi^{-} \ell^{+} {\nu}_{\ell} \), \( D^{*+}_{s} \to K^{0} \ell^{+} {\nu}_{\ell} \), \( D^{*0} \to \rho^{-} \ell^{+} {\nu}_{\ell} \) and \( B_c^{*+} \to B^{*0} \ell^{+} {\nu}_{\ell} \). A second category, the \( b \to c \ell^{-} \bar{\nu}_{\ell} \) transition, features decays like \( B_c^{*+} \to \psi \,\ell^{-} \bar{\nu}_{\ell} \), \( B^{*-} \to D^{*0} \ell^{-} \bar{\nu}_{\ell} \), \( B^{*0} \to D^{*+}(D^{+}) \ell^{-} \bar{\nu}_{\ell} \), and \( B^{*0}_{s} \to D^{*+}_{s} (D^{+}_{s}) \ell^{-} \bar{\nu}_{\ell} \). Meanwhile, the \( c \to s  \ell^{+} {\nu}_{\ell}\) mode encompasses processes such as \( \psi \to D^{*-}_{s}(D^{-}_{s})  \ell^{+} {\nu}_{\ell} \) and \( D^{*0} \to K^{*-} (K^{-})  \ell^{+} {\nu}_{\ell} \). Finally, the \( b \to u \ell^{-} \bar{\nu}_{\ell}\) channel includes decays like \( B_c^{*-} \to D^{*0} \ell^{-} \bar{\nu}_{\ell} \), \( \bar{B_{s}}^{*0} \to K^{*+} \ell^{-} \bar{\nu}_{\ell} \), and \( \bar{B}^{*0} \to \rho^{+} \ell^{-} \bar{\nu}_{\ell} \).
For the semi-leptonic decays  \( V \to V (P)\ell^{+} \nu_{\ell}(\ell^{-} \bar{\nu}_{\nu}) \) the effective weak Hamiltonian is given by:
\begin{eqnarray}\label{eq.efh}
\mathcal{H}_{eff}(Q\to Q'\ell \bar{\nu})= \frac{G_{f}}{\sqrt{2}}V_{QQ'}\,\big( \bar{Q}\,\gamma_{\mu}(1-\gamma_{5})\,Q'\,\big)\,\big( \ell^{+}(\ell^{-})\,
\gamma^{\mu}(1-\gamma_{5})\,\nu_{\ell}(\bar{\nu}_{\ell})\big),
\end{eqnarray}

where \( G_F \) is the Fermi constant, and \( V_{QQ'} \) denotes the CKM matrix element for the quark transition \( Q \to Q' \).
To compute the rate of a semi-leptonic decay \( V \to V (P)\ell^{+} \nu_{\ell}(\ell^{-} \bar{\nu}_{\nu}) \), the essential component in this analysis is the hadronic matrix element of the weak current, evaluated between the initial and final meson states. This matrix element is conventionally expressed in terms of Lorentz-invariant form factors. For \( V \to V \) transition, we define:
\begin{eqnarray}\label{eq.deffform}
\langle V(p_{2},\varepsilon_{2})|\bar{Q}\,\gamma_{\mu}(1-\gamma_{5})\,Q'|V(p_{1}, \varepsilon_{1})\rangle&=
(\varepsilon_{1}.\varepsilon_{2}^{*})\left[-(p_{1}+p_{2})_{\mu}\,V_{1}(q^2)+q_{\mu}\,V_{2}(q^2)\right]-(\varepsilon_{1}.q)
\,\varepsilon_{2\mu}^{*}V_{3}(q^2),\label{eq.deffform1}
\end{eqnarray}

where  \( q = p_1 - p_2 \) denotes the momentum transfer and \( q^2 \)  represents its invariant squared mass.
The polarization vectors \( \varepsilon_1 \) and \( \varepsilon_2 \) correspond to the initial and final vector mesons \( V \), respectively. In Eq. (\ref{eq.deffform1}), the dimensionless form factors \( V_1 \), \( V_2 \), and \( V_3 \)  govern the transition amplitude.\\ 
For the  \( V \to P \) transition, we define:  
\begin{eqnarray}
\langle P(p_{2})|\bar{Q}\,\gamma_{\mu}\,(1-\gamma_{5})\,Q'|V(p_{1}, \varepsilon_{1})\rangle&=&\varepsilon_{1\mu}^{*}\,m_{_P}\,m_{_V}\, \,\frac{2\,V(q^2)}{(m_{_P}+m_{_V})}+
i\, \varepsilon_{_1\mu}^{*}\,(m_{_P}+m_{_V}) \,A_{1}(q^2)\nonumber\\
&&-i\,(p_{1}+p_{2})_{\mu}~ (\varepsilon_{_1}^{*}.q)\,
\frac{A_{2}(q^2)}{(m_{_P}+m_{_V})}
 -i\,\,q_{\mu}~(\varepsilon_{_1}^{*}.q)\,\frac{2\,m_{_P}}{q^2} \,\label{eq.deffform2}
 [A_{3}(q^2)-A_{0}(q^2)],
\end{eqnarray}

In Eq. (\ref{eq.deffform2}),
 $V(q^2)$ and
 $A_{i}(q^2)$ with $(i=0,...,3)$ are the transition form factors for 
vector-to-pseudoscalar transition. The form factor $A_{3}(q^2) $ 
  can be expressed as a linear combination of $A_{1}(q^2) $ and $A_{2}(q^2) $
 as:
\begin{eqnarray}\label{eq28}
 A_{3}(q^2)=\frac{(m_{_P}+m_{_V})}{2\,m_{_P}}\, A_{1}(q^2)
 -\frac{(m_{_V}-m_{_P})}{2\,m_{_P}}\, A_{2}(q^2),
\end{eqnarray}

with the condition $A_{0}(0)=A_{3}(0)$.
 
Within the Hard-Wall AdS/QCD framework, we compute the transition form factors for \( V \to V(P) \) transitions using three-point correlation functions involving the initial state, final state, and weak interaction currents.
 These 3-point correlation functions are derived by functionally differentiating the 5D bulk action with respect to their corresponding sources. 
 These sources are defined as the boundary values of the 5D fields carrying the appropriate quantum numbers as \cite{Witten1998, Gubser1998, Grigoryan2007, Abidin2008}:
\begin{eqnarray}
\langle 0 |\mathcal{T}\left.\{ J_{V\perp}^{\mu a} (x) J_{V\perp}^{\nu b}(y)
	J_{V\perp}^{\sigma c}(w)\right.\}	| 0 \rangle&=&-\frac{\ \delta^{3} \mathcal{S}\rm{(VVV)} \qquad }
	{\delta V_{\perp\mu}^{0a}(x) \,
	\delta V_{\perp\nu}^{0b}(y) \, \delta V_{\perp\sigma}^{0c}(w)},~~~~~ \label{3p1} \\
\nonumber\\
\langle 0 |\mathcal{T}\left.\{ J_{V\perp}^{\mu a} (x) J_{A\perp}^{\nu b}(y)
	J_{V\perp}^{\sigma c}(w)\right.\}	| 0 \rangle&=&-\frac{\ \delta^{3} \mathcal{S}\rm{(VAV)} \qquad }
	{\delta V_{\perp\mu}^{0a}(x) \,
	\delta A_{\perp\nu}^{0b}(y) \, \delta V_{\perp\sigma}^{0c}(w)},~~~~~ \label{3p2} \\
\nonumber\\
\langle 0 |\mathcal{T}\left.\{ J_{A\parallel}^{\alpha a}(x) J_{V\perp}^{\mu b}(y)  J_{V\perp}^{\nu c} (w)
	\right.\}	| 0 \rangle&=&-\frac{\ \delta^{3} \mathcal{S}\rm{(PVV)} \qquad }
	{\delta A_{\parallel\alpha}^{0a}(x) \delta V_{\perp\mu}^{0b}(y) \,
	\delta V_{\perp\nu}^{0c}(w) \, },~~~~~ \label{3p3} \\
\nonumber\\
\langle 0 |\mathcal{T}\left.\{ J_{A\parallel}^{\alpha a}(x) J_{A\perp}^{\mu b}(y)  J_{V\perp}^{\nu c} (w)
	\right.\}	| 0 \rangle&=&-\frac{\ \delta^{3} \mathcal{S}\rm{(PAV)} \qquad }
	{\delta A_{\parallel\alpha}^{0a}(x) \delta A_{\perp\mu}^{0b}(y) \,
	\delta V_{\perp\nu}^{0c}(w) \, },~~~~~ \label{3p4}
\end{eqnarray}

In Eqs. (\ref{3p1}-\ref{3p4}), the quantity $\mathcal{S}\rm{(M_{1}, M_{2}, M_{3})}$ represents the relevant terms in the five-dimensional action that incorporate the fields corresponding to the states $\rm{(M_{1}, M_{2}, M_{3})}$.
 Using Eqs. (\ref{3p1}-\ref{3p4}), we obtain:
\begin{eqnarray}
\langle 0 |\mathcal{T}\left.\{ J_{V\perp}^{\mu a} (x) J_{(V-A)\perp}^{\nu b}(y)
	J_{V\perp}^{\sigma c}(w)\right.\}	| 0 \rangle&&=-\frac{\ \delta^{3} \mathcal{S}\rm{(VVV)} \qquad }
	{\delta V_{\perp\mu}^{0a}(x) \,
	\delta V_{\perp\nu}^{0b}(y) \, \delta V_{\perp\sigma}^{0c}(w)}+\frac{\ \delta^{3} \mathcal{S}\rm{(VAV)} \qquad }
	{\delta V_{\perp\mu}^{0a}(x) \,
	\delta A_{\perp\nu}^{0b}(y) \, \delta V_{\perp\sigma}^{0c}(w)},\label{3p12} \\
\langle 0 |\mathcal{T}\left.\{J_{A\parallel}^{\alpha a}(w) J_{(V-A)\perp}^{\mu b}(y)
	 J_{V\perp}^{\nu c}(w)\right.\}	| 0 \rangle&&=-\frac{\ \delta^{3} \mathcal{S}\rm{(PVV)} \qquad }
	{\delta A_{\parallel\alpha}^{0a}(x) \delta V_{\perp\mu}^{0b}(y) \,
	\delta V_{\perp\nu}^{0c}(w) \, }+\frac{\ \delta^{3} \mathcal{S}\rm{(PAV)} \qquad }
	{\delta A_{\parallel\alpha}^{0a}(x) \delta A_{\perp\mu}^{0b}(y) \,
	\delta V_{\perp\nu}^{0c}(w) \, }.~~~~~ \label{3p22}
\end{eqnarray}

To determine the weak form factors, we insert two complete sets of intermediate states with quantum numbers matching those of the initial and final meson currents and  use the following definitions:
\begin{eqnarray}
\langle0|J_{V\perp}^{\tau a}|V^{a'}(p, \varepsilon_{_V})\rangle&=&f_{_V}\,\varepsilon_{_V}^{\tau}\,\delta^{aa'},\label{decayv}
\\
\langle0|J_{A\parallel}^{\alpha d}|\phi^{d'}(p)\rangle&=&i f_{_P}^d\, p^\alpha \delta^{dd'},\label{decayp}
\end{eqnarray}
where, \(f_{_V}\), and \(f_{_P}\) denote the decay constants of the vector, and pseudoscalar mesons, respectively. We now  arrive at:
\begin{eqnarray}
\langle V(p_{2},\varepsilon_{2})|\bar{Q}\,\gamma_{\mu}(1-\gamma_{5})\,Q'|V(p_{1}, \varepsilon_{1})\rangle&=&\,
\Lambda[(V, p_{2}); (V, p_{1})]\,\varepsilon_{_2\nu}^{*}\,\varepsilon_{_1\sigma}\nonumber\\
&\times&\hat{\mathcal{I}}\Bigg(-\frac{\ \delta^{3} \mathcal{S}\rm{(VVV)} \qquad }
	{\delta V_{\perp\nu}^{0a}(x) \,
	\delta V_{\perp\mu}^{0b}(0) \, \delta V_{\perp\sigma}^{0c}(w)}+\frac{\ \delta^{3} \mathcal{S}\rm{(VAV)} \qquad }
	{\delta V_{\perp\nu}^{0a}(x) \,
	\delta A_{\perp\mu}^{0b}(0) \, \delta V_{\perp\sigma}^{0c}(w)}\Bigg), \label{f1}\\
\langle P(p_{2})|\bar{Q}\,\gamma_{\mu}\,(1-\gamma_{5})\,Q'|V(p_{1}, \varepsilon_{1})\rangle&=&\,
\Lambda[(P, p_{2}); (V, p_{1})]\,\varepsilon_{_1\nu}\,\Omega_{\alpha}(p_{2})\nonumber\\
&\times&\hat{\mathcal{I}}\Bigg(-\frac{\ \delta^{3} \mathcal{S}\rm{(PVV)} \qquad }
	{\delta A_{\parallel\alpha}^{0a}(x) \delta V_{\perp\mu}^{0b}(y) \,
	\delta V_{\perp\nu}^{0c}(w) \, }+\frac{\ \delta^{3} \mathcal{S}\rm{(PAV)} \qquad }
	{\delta A_{\parallel\alpha}^{0a}(x) \delta A_{\perp\mu}^{0b}(y) \,
	\delta V_{\perp\nu}^{0c}(w) \, }\Bigg), \label{f2}
\end{eqnarray}

where the double integration operator \(\hat{\mathcal{I}}\) is defined as:
\begin{eqnarray}
\hat{\mathcal{I}} = \int d^4x \, d^4w \, e^{ip_{1} x - ip_{3} w}.
\end{eqnarray}

which is applied after performing the Fourier transform.
 Furthermore, we introduce the momentum-space propagator \(\Omega_{\alpha}(p_{k}) = \frac{p_{k\alpha}}{p_{k}^{2}}\), and the expression for \(\Lambda[(\mathcal{O}_{2}, p_{2}); (\mathcal{O}_{1}, p_{1})]\) is given by:
\[
\Lambda[(\mathcal{O}_{2}, p_{2}); (\mathcal{O}_{1}, p_{1})]= \frac{(p_{1}^2 - m^{2}_{\mathcal{O}_{1}})}{f_{\mathcal{O}_{1}}} \, \frac{(p_{2}^2 - m^{2}_{\mathcal{O}_{2}})}{f_{\mathcal{O}_{2}}},
\]

in our notation. Here, $f_{\mathcal{O}_{i}}$ denotes the decay constant for the $i^{th}$ meson. In our final result, we will consider the on-shell limit  $(p_{1}^2, p_{2}^2) \to (m^{2}_{\mathcal{O}_{1}}, m^{2}_{\mathcal{O}_{2}})$. By taking this on-shell limit, we effectively perform  a Lehmann-Symanzik-Zimmermann (LSZ) reduction that isolates the double pole corresponding to the ground-state mesons. Consequently, the excited and continuum states do not explicitly appear in the final form factor expressions, as we have mathematically projected out the ground-state matrix elements. It is important to emphasize that their dynamical effects are not discarded or approximated away; they are implicitly resummed and included through the exact bulk-to-boundary propagators (e.g., $\mathcal{V}^b(q^2, z)$), which are the exact continuous solutions to the 5D equations of motion and contain the infinite tower of AdS modes (see Eqs. \ref{eq.gra}-\ref{eq.grpit}). This is fundamentally different from traditional QCD sum rules, as our approach does not require a ground-state saturation approximation or a separate continuum subtraction.\\
We now require $\mathcal{S}\rm{(VVV)}$, $\mathcal{S}\rm{(VAV)}$, $\mathcal{S}\rm{(PVV)}$, and $\mathcal{S}\rm{(PAV)}$. 
The relevant actions are given by \cite{Momeni2021}:
\begin{eqnarray}
\mathcal{S}\rm{(VVV)}&=&\int d^5x\bigg( \frac{f^{abc}}{g_{5}^2\,z}\left.[V_{\mu\nu}^{a}\,V^{\nu b}\,V^{\nu c}\right.]\bigg),\label{svvv}\\
\mathcal{S}\rm{(PVV)}&=&\int d^5x\bigg(
\frac{k^{abc}}{z^3}\left.[\pi^{a}\,V^{b\mu}\,V_{\mu}^{c}+\pi^{a}\,V^{bz}\,V_{z}^{c}\right.]\bigg) \label{spvv},\\
\mathcal{S}\rm{(PAV)}&=&\int d^5x\bigg( \frac{f^{abc}}{2\,g_{5}^2\,z}\left.[\partial^{\mu}\phi^{a}\,A^{\nu b}\,V_{\mu\nu}^{c}\right.] +
\frac{h^{abc}}{z^3}\left.[\pi^{a}\,A_{\mu}^{b}\,V^{c\mu}\right.] +\frac{g^{abc}}{z^3}\left.[\pi^{a}\,A_{\mu}^{b}\,V^{c\mu}\,
\right.] \bigg)\label{spav}.
\end{eqnarray}

The contribution to $\mathcal{S}({\text{VAV}})$ from the left-handed gauge field term ($L_{MN}L^{MN}$) has an opposite sign compared to that from the right-handed term ($R_{MN}R^{MN}$). Consequently, these terms cancel exactly, yielding $\mathcal{S}({\text{VAV}}) = 0$.\\
In this context, \( f^{abc} \) denotes the structure constants of \( \text{SU}(5) \), and the terms containing this factor originate from the gauge sector of the leading-order action. The following definitions are also relevant to Eqs. (\ref{svvv}, \ref{spvv}, and \ref{spav}): 
\begin{eqnarray}
k^{abc} &=& -2i \, \text{Tr} \left( [t^a, X_0] [t^b, \{t^c, X_0\}] \right), \\
h^{abc} &=& -2i \, \text{Tr} \left( [t^a, X_0] \{t^b, \{t^c, X_0\}\} \right), \\
g^{abc} &=& -2i \, \text{Tr} \left( \{t^a, X_0\} [t^b, \{t^c, X_0\}] \right).
\end{eqnarray}

The structure constants \( f^{abc} \) are adopted from \cite{Mahmoud2013}. For the numerical computations, the coefficients \( k^{abc} \), \( h^{abc} \), and \( g^{abc} \) are taken from \cite{Momeni2022}.  
Using the Fourier transforms defined in \cite{Grigoryan20072,Abidin20083}, as:
\begin{align}
\phi^a(p,z) &= \phi^a(p^2,z) \frac{i p^\alpha}{p^2} A_{\parallel\alpha}^{0a}(p), \\
\pi^a(p,z) &= \pi^a(p^2,z) \frac{i p^\alpha}{p^2} A_{\parallel\alpha}^{0a}(p), \\
A_{\perp \mu}^a(q,z) &= \mathcal{A}^a(q^2,z) A_{\perp \mu}^{0a}(q), \\
V_{\perp \mu}^b(q,z) &= \mathcal{V}^b(q^2,z) V_{\perp \mu}^{0b}(q), \\
V_z^b(q,z) &= -\partial_z \tilde\pi^b(q^2,z) \frac{i q^\alpha}{q^2} V_{\parallel \alpha}^{0b}(q),\\
\partial^\mu &=\to -i \left( \text{relevant momentum} \right)^\mu,
\end{align}

the form factors for \( V \to V \, \ell \, \bar{\nu} \) can be derived as follows:
\begin{eqnarray}
V_{1}(q^2)&=& \int_{0}^{z_{0}} dz\,\frac{f^{abc}}{z} \psi_{V}^{a}(z)\,\mathcal{V}^{b}(q^2, z) \,\psi_{V}^{c},\label{eq.V1}\\
V_{2}(q^2)&=& \int_{0}^{z_{0}} dz\,\frac{f^{abc}}{2\,z} \psi_{V}^{a}(z)\,\mathcal{V}^{b}(q^2, z) \,\psi_{V}^{c},\label{eq.V2}\\
V_{3}(q^2)&=& -\int_{0}^{z_{0}} dz\,\frac{f^{abc}}{z} \psi_{V}^{a}(z)\,\mathcal{V}^{b}(q^2, z) \,\psi_{V}^{c}.\label{eq.V3}
\end{eqnarray}

Moreover, the transition form factors governing the \( V \to P \, \ell \, \bar{\nu} \) decay process are obtained as:
\begin{eqnarray}
V(q^2)&=& \frac{(m_{_P}+m_{_V})}{2}\,\frac{g_5^2}{m_{_P}\,m_{_V}}\int_{0}^{z_{0}} dz\,\frac{k^{abc}}{z^3} \pi^{a}(z)\,\mathcal{V}^{b}(q^2, z) \,\psi_{V}^{c},\label{eq.V}\\
A_{1}(q^2)&=&\frac{(q^2+m_{_V}^2-m_{_P}^2)}{(m_{_P}+m_{_V})}\int_{0}^{z_{0}} dz\,\frac{f^{abc}}{4\,z}\, \phi^{a}(z)\,\mathcal{A}^{b}(q^2, z) \,\psi_{V}^{c}\nonumber\label{eq.A1}\\
&+&\frac{g_{5}^2}{(m_{_P}+m_{_V})}\int_{0}^{z_{0}} dz\,\frac{(h^{abc}+g^{abc})}{z^3}\, \pi^{a}(z)\,\mathcal{A}^{b}(q^2, z) \,\psi_{V}^{c}\\
A_{2}(q^2)&=&  \frac{(m_{_P}+m_{_V})}{4}\int_{0}^{z_{0}} dz\,\frac{f^{abc}}{z} \phi^{a}(z)\,\mathcal{A}^{b}(q^2, z) \,\psi_{V}^{c},\label{eq.A2}\\
A_{0}(q^2)&=&\frac{g_{5}^{2}\,(3\,m_{_P}^{2}-q^2-\,m_{_V}^{2})}{4\,m_{_P}}\,\int_{0}^{z_{0}}\,dz\,
\frac{f^{abc}}{4\,z}\,\phi^{a}(z)\,\,{\cal A}^b(q^2,z)\,\psi_{_V}^{c}(z)\nonumber\\
\nonumber\\
&+&\frac{g_{5}^{2}}{4\,m_{_P}}\int_{0}^{z_{0}}\,dz\,
\frac{(h^{abc}+g^{abc})}{z^3}\, \pi^{a}(z)\,\mathcal{A}^{b}(q^2, z) \,\psi_{V}^{c}\label{eq.A0},
\end{eqnarray}

It is worth emphasizing that unlike traditional QCD sum rules, the Hard-Wall AdS/QCD framework does not require a ground-state saturation approximation or a separate continuum subtraction. Although the propagators are formally expressed as a sum over poles (Eqs. \ref{eq.gra}-\ref{eq.grpit}), the form factors in Eqs. (\ref{eq.V1}-\ref{eq.A0}) are computed using the exact, continuous solutions to the 5D equations of motion. By retaining the full bulk-to-boundary propagators, the overlap integrals inherently resum the contributions of all radially excited states. Furthermore, the infinite tower of these discrete AdS modes naturally reproduces the perturbative QCD continuum above the confinement scale set by $z_0$.

\section{Numerical analysis}\label{sec.3}
With the analytic expressions for the form factors of the vector-to-vector (pseudoscalar)  transitions derived in the previous section, we now specify the input parameters to estimate phenomenologically relevant quantities. 
To evaluate the form factors for \( V \to V(P) \) decays, we must first establish the values of the parameters \( z_0 \), quark masses (\( m_u, m_d, m_s, m_c, m_b \)), and quark condensates (\( \sigma_u, \sigma_d, \sigma_s, \sigma_c, \sigma_b \)). These parameters, along with their associated uncertainties, were previously determined in Ref. \cite{Momeni2023} through a global fit to the experimental masses of the mesons \(\rho^{0}\), \(\rho^{-}\), \(a_{1}^{-}\), \(\pi^{0}\), \(\pi^{-}\), \(K^{-}\), \(K^{*-}\), \(D^{-}\), \(D^{*-}\), \( B^{-}\), and \(B^{*0}\).  
The best-fit values are as follows: the inverse mass scale is found to be \( z_0^{-1} = (323 \pm 1) \, \text{MeV} \), the quark masses (in MeV) are \( m_u = 8.5 \pm 2.5 \), \( m_d = 12.36 \pm 2.45 \), \( m_s = 195.31 \pm 5.89 \), \( m_c = 1590.56 \pm 8.42 \), and  $m_{b}=(4120 \pm 5)$  while the quark condensates (in \(\text{MeV}^3\)) are \( \sigma_u = (173.65 \pm 2.21)^3 \), \( \sigma_d = (177.42 \pm 3.15)^3 \), \( \sigma_s = (226.20 \pm 3.72)^3 \), \( \sigma_c = (310.35 \pm 5.65)^3 \), and $\sigma_{b}=(420 \pm 1.12)^3$.

We note a crucial distinction regarding the quark masses utilized in this work: they are not the familiar perturbative masses (i.e., pole or $\overline{\text{MS}}$
 masses). Because our analysis is grounded in the Hard-Wall AdS/QCD model, we employ effective, constituent-like masses inherent to this holographic framework. These bulk mass parameters naturally encapsulate non-perturbative dynamics, such as chiral symmetry breaking and confinement effects introduced by the IR cutoff $z_0$-which are absent in perturbative definitions. Practically, these masses are treated as free model parameters, fixed by fitting the predicted hadron spectra to experimental observations. As a result, they define an AdS/QCD-specific mass scheme that is distinct from perturbative schemes; bridging the two would require non-trivial matching procedures beyond the scope of the current analysis.
 
We note that the fitted values of $\sigma_q$ in this model act as effective phenomenological parameters rather than the fundamental QCD quark condensates. Due to the rescaling parameter $\zeta$ and the absence of the QCD running coupling in the pure AdS hard-wall background, the values of $\sigma_q$ deviate from standard LQCD predictions \cite{McNeile2013}. However, as effective constituent-like parameters fitted to reproduce the hadronic spectra, they successfully yield physical observables such as decay constants and masses that are consistent with LQCD results.

In Refs. \cite{Momeni2021, Momeni2023}, the wave functions, masses, and decay constants of  mesons were systematically investigated within the framework of Hard-Wall AdS/QCD with five active flavors ($N_{f}=5$).  For the numerical computation of the form factors discussed in Section \ref{sec.2}, the masses of the initial-state \(V\) mesons and the final-state \(V/P\) mesons are essential inputs. These masses are summarized in Table \ref{massr}. 
The mass values for \(D^{*-}\), \(K^{-}\), \(D^{-}\), \(B^{*-}\), \(K^{*-}\), \(\pi^{-}\), and \(\rho^{-}\) are taken from experimental data in Ref. \cite{pdg}, while our model in Refs. \cite{Momeni2021, Momeni2023} predicts the masses of \(\psi\), \(B_c^{*-}\), \(B_s^{*0}\), \(D^{*0}\), \(D_s^{*-}\), \(D_s^{-}\), and \(K^{0}\). 
 
\begin{table}[th]
\caption{Masses of the initial-state vector (\(V\)) and final-state vector/pseudoscalar (\(V/P\)) mesons used in the numerical computation of the form factors (Section \ref{sec.2}). Experimental values for \(D^{*-}\), \(K^{-}\), \(D^{-}\), \(B^{*-}\), \(K^{*-}\), \(\pi^{-}\), and \(\rho^{-}\) are taken from Ref. \cite{pdg}, while the masses of \(\psi\), \(B_c^{*-}\), \(B_s^{*0}\), \(D^{*0}\), \(D_s^{*-}\), \(D_s^{-}\), and \(K^{0}\) are theoretical predictions from the Hard-Wall AdS/QCD model in Refs. \cite{Momeni2021}, \cite{Momeni2023}.}
\label{massr}
\begin{tabular}{|c||c|c||c|c|}
\hline
Meson  &Mass ~(MeV)&Meson  &Mass ~(MeV)\\
\hline
&&&\\
$B^{*-}$&$5324.71 \pm 0.21 $&$D^{-}$&$1869.65\pm 0.05$\\
&&&\\
$B_s^{{*0}}$&$5398.32\pm 8.16$&$D_{s}^{-}$&$1972.63\pm 2.37$\\
&&&\\
$B_c^{*-}$&$6510.14\pm 13.32$&$K^{0}$&$499.21\pm 1.82$\\
&&&\\
$\psi$&$3095.20\pm 0.15$&$K^{*-}$&$891.66 \pm 0.26$\\
&&&\\
$D^{*0}$&$2005.53\pm 6.65$&$\rho^-$&$775.40\pm 0.34$\\
&&&\\
$D_{s}^{*-}$&$2112.90\pm 9.42$&$K^{-}$&$493.67\pm 0.01$\\
&&&\\
$D^{*-}$&$2010.26\pm 0.05$&$\pi^{-}$&$139.75\pm 0.00$\\
&&&\\
\hline
\end{tabular}
\end{table}

We investigate the form factors for the previously discussed semileptonic decays as a function of \( q^2 \).  Our predictions at \(q^2 = 0\) are summarized in two tables: the \(V \to P \ell^{+} \nu_{\ell} (\ell^{-} \bar{\nu}_{\ell})\) form factors, which cover a wide spectrum of heavy-to-light transitions, are given in Table \ref{valqzero}; the corresponding results for \(V \to V \ell^{+} \nu_{\ell} (\ell^{-} \bar{\nu}_{\ell})\) decays are provided in Table \ref{valqzero2}. Both tables, generated within the Hard-Wall AdS/QCD model, list the central values and their associated theoretical uncertainty regions. The largest contribution to the uncertainties in the form factors comes from the error in the quark condensate (approximately $(85-90)\%$), while the remaining uncertainty originates from the masses.

\begin{table}
\centering
\caption{The form factors for the semileptonic decays \(V \to P \ell^{+} \nu_{\ell} (\ell^{-} \bar{\nu}_{\ell})\), where \(V = (\psi, D^{*0},  D^{*+}_s, B^{*0}, B^{*0}_s)\) and \(P = (D^{\pm}, D^{\pm}_s, K^{-}, K^{0}, \pi^{-})\), calculated at zero momentum transfer \(q^2 = 0\) within the Hard-Wall AdS/QCD model. The associated uncertainty regions are also shown. }\label{valqzero}
\begin{ruledtabular}
\begin{tabular}{|c|c|c|c|c|c|c|c|}
Form factor  &Value at $q^2=0$ &Form factor&Value at $q^2=0$&Form factor&Value at $q^2=0$  &Form factor&Value at $q^2=0$  \\
\hline
\hline
&&&&&&&\\
$V^{\psi \to D^{-}}$&$1.21^{+0.12}_{-0.14}$&$A_{3}^{\psi \to D_{s}^{-}}$&$0.85^{+0.13}_{-0.10}$&$A_{2}^{D^{*+}_{s} \to K^{0}}$&$0.51 ^{+0.08}_{-0.06}$&$A_{1}^{  B_s^{*0} \to D_s^{+} }$&$0.65^{+0.15}_{-0.14}$\\
&&&&&&&\\
$A_{1}^{\psi \to D^{-}}$&$0.64^{+0.09}_{-0.08}$&$V^{D^{*0} \to K^{-}}$&$1.05^{+0.12}_{-0.10}$&$A_3^{D^{*+}_{s} \to K^{0}}$&$0.88^{+0.10}_{-0.13}$&$A_{2}^{  B_s^{*0} \to D_s^{+} }$&$0.57^{+0.12}_{-0.11}$\\
&&&&&&&\\
$A_{2}^{\psi \to D^{-}}$&$0.42^{+0.10}_{-0.08}$&$A_{1}^{D^{*0} \to K^{-}}$&$0.72^{+0.08}_{-0.07}$&$V^{D^{*0} \to \pi^{-}}$&$0.85^{+0.09}_{-0.08}$&$A_{3}^{  B_s^{*0} \to D_s^{+} }$&$0.72^{+0.16}_{-0.14}$\\
&&&&&&&\\
$A_{3}^{\psi \to D^{-}}$&$0.71^{+0.08}_{-0.06}$&$A_{2}^{D^{*0} \to K^{-}}$&$0.55^{+0.07}_{-0.08}$&$A_1^{D^{*0} \to \pi^{-}}$&$0.69^{+0.12}_{-0.10}$&$V^{B^{*0} \to D^{+} }$&$0.82^{+0.21}_{-0.18}$\\
&&&&&&&\\
$V^{\psi \to D_{s}^{-}}$&$1.32^{+0.18}_{-0.14}$&$A_{3}^{D^{*0} \to K^{-}}$&$0.98^{+0.12}_{-0.11}$&$A_{2}^{D^{*0} \to \pi^{-}}$&$0.73^{+0.13}_{-0.10}$&$A_{1}^{B^{*0} \to D^{+}}$&$0.69^{+0.15}_{-0.18}$\\
&&&&&&&\\
$A_{1}^{\psi \to D_{s}^{-}}$&$0.78^{+0.14}_{-0.13}$&$V^{D^{*+}_{s} \to K^{0}}$&$1.09^{+0.10}_{-0.08}$&$A_{3}^{D^{*0} \to \pi^{-}}$&$0.43^{+0.12}_{-0.10}$&$A_{2}^{B^{*0} \to D^{+}}$&$0.62^{+0.13}_{-0.10}$\\
&&&&&&&\\
$A_{2}^{\psi \to D_{s}^{-}}$&$0.52^{+0.12}_{-0.10}$&$A_{1}^{D^{*+}_{s} \to K^{0}}$&$0.66^{+0.08}_{-0.07}$&$V^{ B_s^{*0} \to D_s^{+}}$&$0.78^{+0.19}_{-0.17}$ &$A_{3}^{B^{*0} \to D^{+}}$&$0.75^{+0.14}_{-0.13}$\\
&&&&&&&\\
\end{tabular}
\end{ruledtabular}
\end{table}
\begin{table}
\centering
\caption{The form factors for the semileptonic decays \(V \to V \), calculated at zero momentum transfer (\(q^2 = 0\)) within the Hard-Wall AdS/QCD model, along with their uncertainty regions. The initial vector meson \(V_i\) is chosen from \(\{\psi, B^{*-}, B^{*0}_{s}, \bar{B}^{*0}_{s}, B^{*0},\bar{B}^{*0}, B^{*-}_{c}, D^{*0}\}\), and the final vector meson \(V_f\) from \(\{ D^{*\pm}, D^{*\pm}_{s}, D^{*0}, B^{*0}, \psi, K^{*\pm}, \rho^{\pm}\}\).}\label{valqzero2}
\begin{ruledtabular}
\begin{tabular}{|c|c|c|c|c|c|c|c|}
Form factor  &Value at $q^2=0$ &Form factor&Value at $q^2=0$&Form factor&Value at $q^2=0$  &Form factor&Value at $q^2=0$  \\
\hline
\hline
 & &&&&  &&\\ 
$V_{1}^{\psi \to D^{*-}}$&$0.67^{+0.10}_{-0.11}$&$V_1^{B^{*0}_{s} \to D^{*+}_{s}}$&$0.98^{+0.18}_{-0.14}$&$V_{1}^{B^{*+}_{c} \to \psi}$&$0.60^{+0.12}_{-0.14}$&$V_{1}^{\bar{B}^{*0} \to \rho^{+} }$&$0.30^{+0.12}_{-0.14}$\\
 & &&&&  &&\\ 
$V_{2}^{\psi \to D^{*-}}$&$0.33^{+0.05}_{-0.05}$&$V_2^{B^{*0}_{s} \to D^{*+}_{s}}$&$0.49^{+0.09}_{-0.07}$&$V_{2}^{B^{*+}_{c} \to \psi}$&$0.30^{+0.06}_{-0.07}$&$V_{2}^{\bar{B}^{*0} \to \rho^{+} }$&$0.15^{+0.06}_{-0.07}$\\
 & &&&&  &&\\ 
$V_{3}^{\psi \to D^{*-}}$&$-0.67^{+0.10}_{-0.11}$&$V_3^{B^{*0}_{s} \to D^{*+}_{s}}$&$-0.98^{+0.18}_{-0.14}$&$V_{3}^{B^{*+}_{c} \to \psi}$&$-0.60^{+0.12}_{-0.14}$&$V_{3}^{\bar{B}^{*0} \to \rho^{+} }$&$-0.30^{+0.10}_{-0.12}$\\
 & &&&&  &&\\ 
$V_{1}^{\psi \to D^{*-}_{s}}$&$0.85^{+0.12}_{-0.13}$&$V_{1}^{B^{*0} \to D^{*+}}$&$0.98^{+0.18}_{-0.14}$&$V_1^{B_c^{*-} \to D^{*0}}$&$0.25^{+0.10}_{-0.08}$&$V_1^{D^{*0} \to \rho^{-} }$&$0.64^{+0.10}_{-0.12}$\\
 & &&&&  &&\\ 
$V_{2}^{\psi \to D^{*-}_{s}}$&$0.44^{+0.06}_{-0.06}$&$V_{2}^{B^{*0} \to D^{*+}}$&$0.49^{+0.09}_{-0.07}$&$V_{2}^{B_c^{*-} \to D^{*0}}$&$0.12^{+0.05}_{-0.04}$&$V_{2}^{D^{*0} \to \rho^{-}}$&$0.32^{+0.05}_{-0.06}$\\
 & &&&&  &&\\ 
$V_{3}^{\psi \to D^{*-}_{s}}$&$-0.85^{+0.12}_{-0.13}$&$V_{3}^{B^{*0} \to D^{*+}}$&$-0.98^{+0.18}_{-0.14}$&$V_{3}^{B_c^{*-} \to D^{*0}}$&$-0.25^{+0.10}_{-0.08}$&$V_{3}^{D^{*0} \to \rho^{-}}$&$-0.64^{+0.10}_{-0.12}$\\
 & &&&&  &&\\ 
$V_{1}^{B^{*-} \to D^{*0}}$&$0.99^{+0.19}_{-0.15}$&$V_{1}^{B^{*+}_{c} \to B^{*0}}$&$1.12^{+0.18}_{-0.20}$&$V_1^{\bar{B}_s^{*0} \to K^{*+}}$&$0.28^{+0.14}_{-0.12}$&$V_{1}^{D^{*0} \to K^{*-}}$&$0.69^{+0.12}_{-0.14}$\\
 & &&&&  &&\\ 
$V_{2}^{B^{*-} \to D^{*0}}$&$0.49^{+0.09}_{-0.07}$&$V_{2}^{B^{*+}_{c} \to B^{*0}}$&$0.56^{+0.09}_{-0.10}$&$V_2^{\bar{B}_s^{*0} \to K^{*+}}$&$0.14^{+0.07}_{-0.06}$&$V_{2}^{D^{*0} \to K^{*-}}$&$0.34^{+0.06}_{-0.07}$\\
 & &&&&  &&\\ 
$V_{3}^{B^{*-} \to D^{*0}}$&$-0.99^{+0.19}_{-0.15}$&$V_{3}^{B^{*+}_{c} \to B^{*0}}$&$-1.12^{+0.18}_{-0.20}$&$V_3^{\bar{B}_s^{*0} \to K^{*+}}$&$-0.28^{+0.14}_{-0.12}$&$V_{3}^{D^{*0} \to K^{*-}}$&$-0.69^{+0.12}_{-0.14}$\\
 & &&&&  &&\\ 
\end{tabular}
\end{ruledtabular}
\end{table}
The form factors for the \(V \to P\,\ell^{+} \nu_{\ell} (\ell^{-} \bar{\nu}_{\ell})\) and \(V \to V\,\ell^{+} \nu_{\ell} (\ell^{-} \bar{\nu}_{\ell})\) decays are calculated using different theoretical frameworks. To ensure a direct comparison of the results, we rescale them according to the form factor definitions in Eqs. (\ref{eq.deffform1}) and (\ref{eq.deffform2}). 

\begin{figure}[th]
\includegraphics[width=8.7cm,height=6cm]{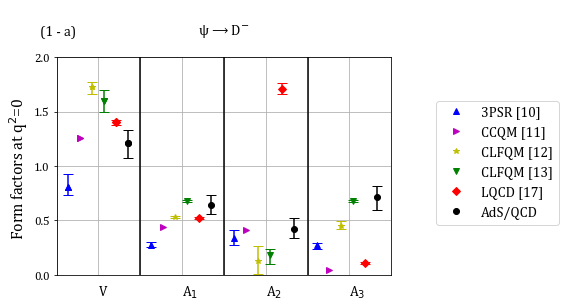}
\includegraphics[width=8.7cm,height=6cm]{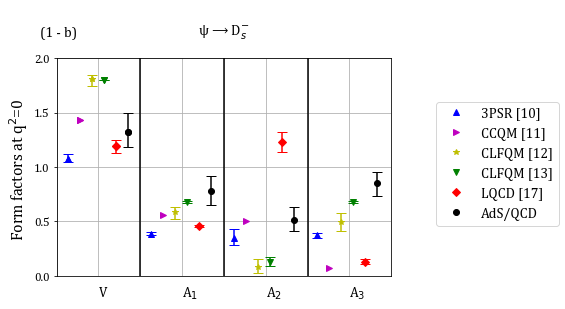}
\caption{Theoretical predictions for the form factors at zero momentum transfer \(q^2=0\). Panel (a) shows the decay \(\psi \to D^{-}\ell\nu_{\ell}\) and panel (b) shows \(\psi \to D_s^{-}\ell\nu_{\ell}\). The results from different models are distinguished by the following markers: triangle-up  for 3PSR \cite{Wang3PSR}, triangle-right for CCQM \cite{Ivanov2015CCQM}, asterisk  for one CLFQM calculation \cite{CLFQM20242}, triangle-down  for another CLFQM calculation \cite{CLFQM20083}, diamond for LQCD \cite{LQCD2008}, and circle  for the present Hard-Wall AdS/QCD results.}\label{C1}
\end{figure}

Figure \ref{C1} shows the form factors at \(q^2=0\) for the decays (a) \(\psi \to D^{-}\,\ell^{+}\,\nu_{\ell}\) and (b) \(\psi \to D_s^{-}\,\ell^{+}\,\nu_{\ell}\), comparing various theoretical approaches.The predictions of these approaches are denoted by distinct markers: triangle-up  for the 3PSR \cite{Wang3PSR}, triangle-right for the  CCQM \cite{Ivanov2015CCQM}. For the CLFQM, we show two variants: an asterisk  represents the calculation from \cite{CLFQM20242}, and a triangle-down  represents the calculation from \cite{CLFQM20083}. Finally, diamond denotes LQCD \cite{LQCD2008}, and circle   our Hard-Wall AdS/QCD results. 

 The results from the CCQM \cite{Ivanov2015CCQM} are presented without uncertainty estimates. 
 As shown in the Fig. \ref{C1}, the form factors \(V\), \(A_1\), and \(A_2\) for the \(\psi \to D^{-}\) channel show relatively good agreement with the CCQM predictions  compared to other models. In contrast, the \(A_3(0)\) form factor in this channel agrees more closely with the CLFQM result \cite{CLFQM20083}. Furthermore, for the \(\psi \to D_s^{-}\) transition, the values of \(A_{1}(0)\) and \(A_{3}(0)\) align better with the CLFQM \cite{CLFQM20083}, while \(V(0)\) and \(A_{2}(0)\) show good agreement with both the CCQM results.
 
\begin{figure}[th]
\includegraphics[width=8.7cm,height=6cm]{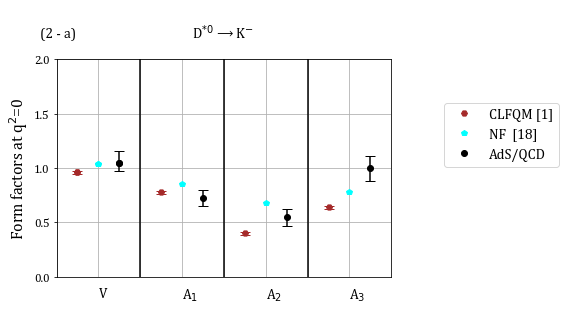}
\includegraphics[width=8.7cm,height=6cm]{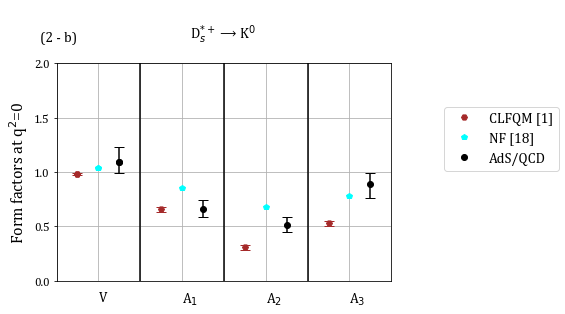}
\includegraphics[width=8.7cm,height=6cm]{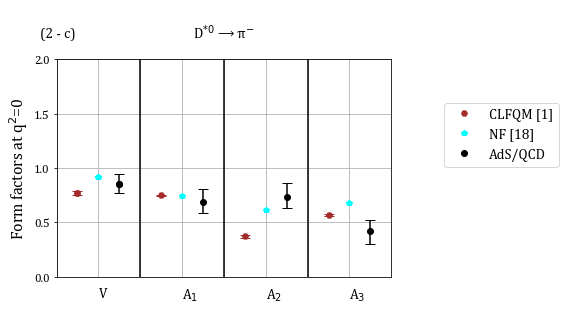}
\caption{Theoretical predictions for the semileptonic decay form factors at zero momentum transfer. The parts show: (a) \(D^{*0} \to K^{-}\), (b) \(D_s^{*+} \to K^{0}\), and (c) \(D^{*0} \to \pi^{-}\). The CLFQM, (hexagon) \cite{CLFQM2024} and the NF (pentagon) \cite{Nfact} are compared with the Hard-Wall AdS/QCD calculation from this work (circle). }\label{C2}
\end{figure}

Figure \ref{C2} presents a complete comparison of the form factors for three transitions: \(D^{*0} \to K^{-} \) in panel (a), \(D_s^{*+} \to K^{0}\) in panel (b), and \(D^{*0} \to \pi^{-}\) in panel (c). The results are distinguished by different markers: a hexagon (CLFQM) \cite{CLFQM20242}, a pentagon (NF) \cite{Nfact}, and a circle (this work). It should be noted that the NF approach does not provide uncertainty estimates. 
From the various panels of Figure \ref{C2}, it is evident that the form factor values for all considered decays show better agreement with the NF approach than with the CLFQM.

\begin{figure}[th]
\includegraphics[width=8.7cm,height=6cm]{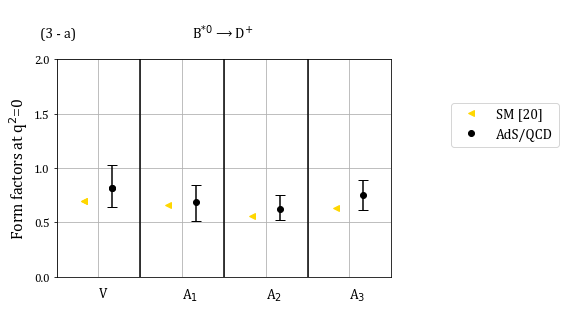}
\includegraphics[width=8.7cm,height=6cm]{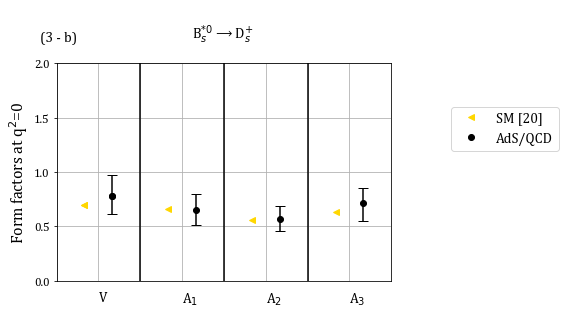}
\caption{Transition form factors for (a) \(B^{*0} \to D^{+}\) and (b) \(B_s^{*0} \to D_s^{+}\) at $q^2=0$. Predictions from the Standard Model (SM) \cite{BtoD2016}, shown with a triangle-left, are presented without uncertainty regions. These are compared with the results of this work, shown with a circle.}\label{C3}
\end{figure}

The form factors for \(B^{*} \to D\,\ell^{+} \nu_{\ell} (\ell^{-} \bar{\nu}_{\ell})\) decays have been evaluated within the SM in \cite{BtoD2016}. Figure \ref{C3} compares these SM predictions \cite{BtoD2016}, which are provided without uncertainty estimates, with our results, using a triangle-left and a circle, respectively. Part (a) displays the \(B^{*0} \to D^{+}\) transition, and part (b) shows the \(B_s^{*0} \to D_s^{+}\) transition. As shown in both panels of this figure, the form factors for the \(B^{*0} \to D^{+}\) and \(B_s^{*0} \to D_s^{+}\) transitions are in good agreement with the SM predictions.

\begin{figure}[th]
\includegraphics[width=8.7cm,height=6cm]{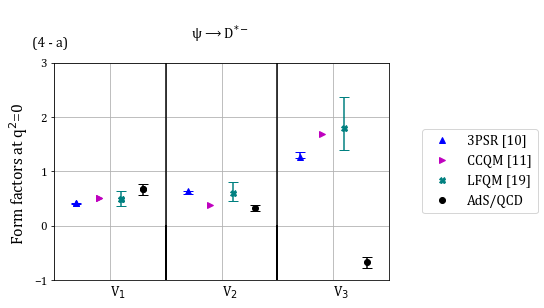}
\includegraphics[width=8.7cm,height=6cm]{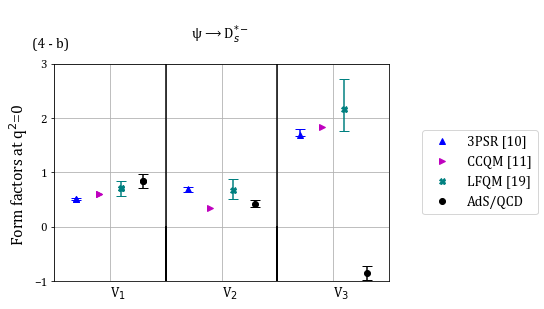}
\caption{Form factors for the (a) \(\psi \to D^{*-}\) and (b) \(\psi \to D_s^{*-}\) transitions at \(q^2=0\). The predictions are distinguished as follows: triangle-up for the 3PSR \cite{Wang3PSR}, triangle-right for the CCQM \cite{Ivanov2015CCQM},  filled-X for the LFQM \cite{LFQM2019}, and circle for this work.}\label{C4}
\end{figure}

We now compare the form factors for \(V \to V\) transitions with predictions from other theoretical models. Figure \ref{C4} shows this comparison for the decays \(\psi \to D^{*-}\) in part (a) and \(\psi \to D_s^{*-}\) in part (b). The results from the 3PSR \cite{Wang3PSR} are denoted by a triangle-up, those from the CCQM \cite{Ivanov2015CCQM} by a triangle-right, and the LFQM \cite{LFQM2019} by a filled-X. From our comparison, we find good agreement between our results and those of the 3PSR, CCQM, and LFQM models for the form factors \(V_1\) and \(V_2\).

\begin{figure}[th]
\includegraphics[width=8.7cm,height=6cm]{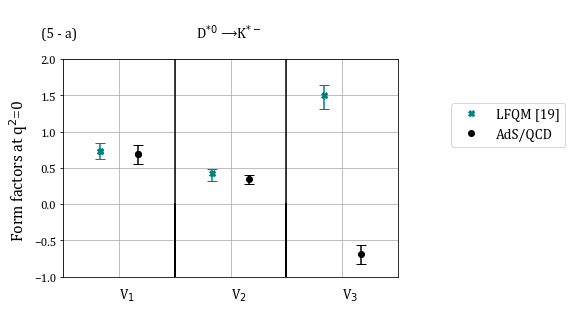}
\includegraphics[width=8.7cm,height=6cm]{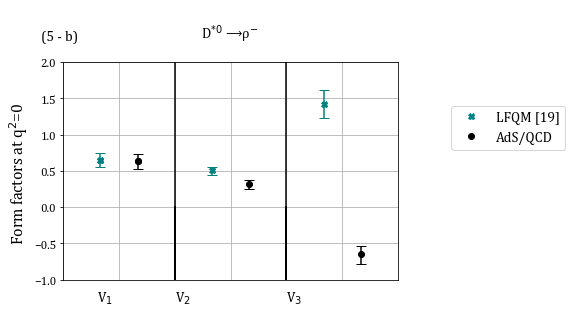}
\caption{ Transition form factors at zero momentum transfer (\(q^2 = 0\)) for (a) \(D^{*0} \to K^{*-}\) and (b) \(D^{*0} \to \rho^{-}\). The predictions are compared with the LFQM \cite{LFQM2019}. The markers for the different theoretical approaches follow the same convention established in Figure \ref{C4}.}\label{C5}
\end{figure}

A comparison of the transition form factors at zero momentum transfer \(q^2 = 0\) is presented in Fig.~\ref{C5}, featuring the decays \(D^{*0} \to K^{*-}\) (part a) and \(D^{*0} \to \rho^{-}\) (part b). The results are compared with predictions from the LFQM \cite{LFQM2019}, using the same marker conventions established in Fig.~\ref{C4}. Our calculations of the \(V_1\) and \(V_2\) form factors agree well with LFQM predictions for both the \(D^{*0} \to K^{*-}\) and \(D^{*0} \to \rho^{-}\) transitions.

\begin{figure}[th]
\includegraphics[width=8.7cm,height=6cm]{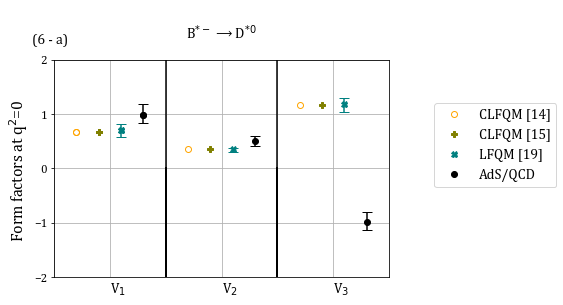}
\includegraphics[width=8.7cm,height=6cm]{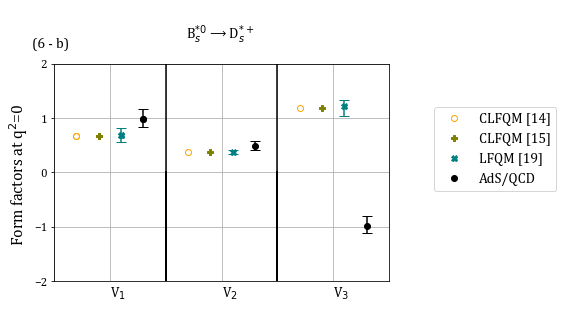}
\includegraphics[width=8.7cm,height=6cm]{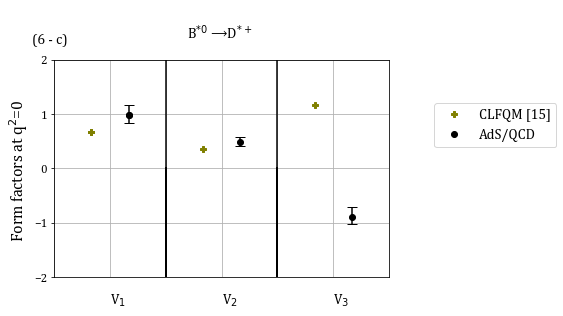}
\includegraphics[width=8.7cm,height=6cm]{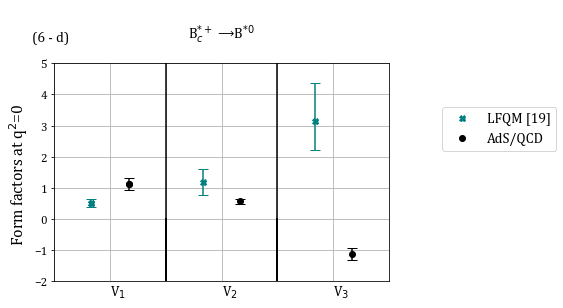}
\includegraphics[width=8.7cm,height=6cm]{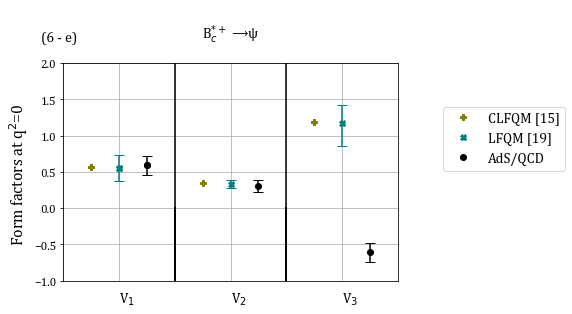}
\includegraphics[width=8.7cm,height=6cm]{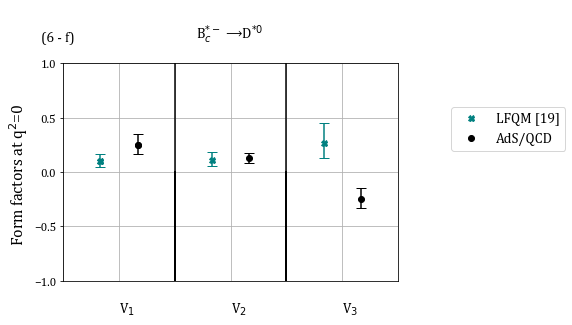}
\includegraphics[width=8.7cm,height=6cm]{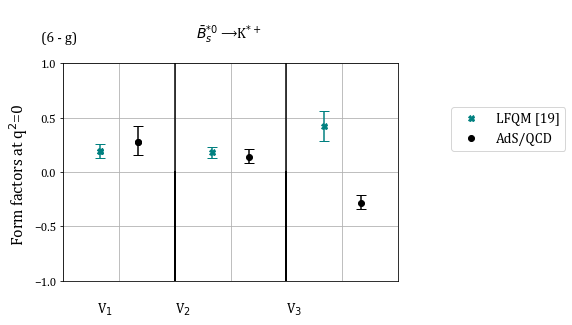}
\includegraphics[width=8.7cm,height=6cm]{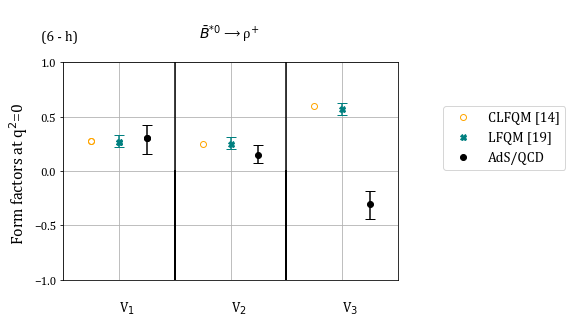}
\caption{ \(B^* \to V\) transition form factors at \(q^2=0\). Panels (a)-(h) show different decay channels. Markers represent: open-circle  (CLFQM \cite{FQCD2019}), filled-plus  (CLFQM \cite{CLFQMVV}), filled-X (LFQM  \cite{LFQM2019}), and circle (this work).}\label{C6}
\end{figure}

The last group in this category involves transitions from vector \(B\) mesons to various vector mesons. A comparison of their form factors at \(q^2=0\) is shown in Figure \ref{C6}, with the following decays displayed in panels (a) to (h): \(B^{*-}\to D^{*0}\), \(B_s^{*0}\to D_s^{*+}\), \(B^{*0}\to D^{*+}\), \(B_c^{*+}\to B^{*0}\), \(B_c^{*+}\to \psi\), \(B_c^{*-}\to D^{*0}\), \(\bar{B}_s^{*0}\to K^{*+}\), and \(\bar{B}^{*0}\to \rho^{+}\). The predictions are distinguished by the markers: open-circle for CLFQM \cite{FQCD2019}, filled-plus for CLFQM \cite{CLFQMVV}, filled-X for LFQM, and  circle for this work.  The results in panels (a) to (h) demonstrate that for the \(V_{1}(0)\) and \(V_{2}(0)\) form factors, our AdS/QCD calculations are consistent with the predictions of both CLFQM models and the LFQM.

\begin{figure}[th]
\includegraphics[width=8.7cm,height=6cm]{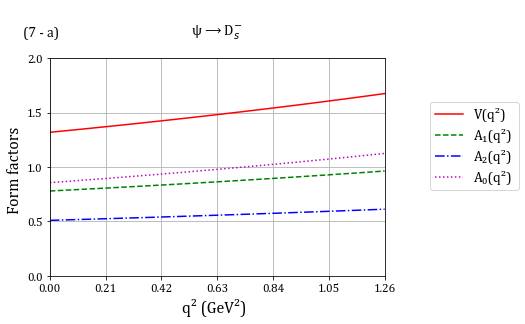}
\includegraphics[width=8.7cm,height=6cm]{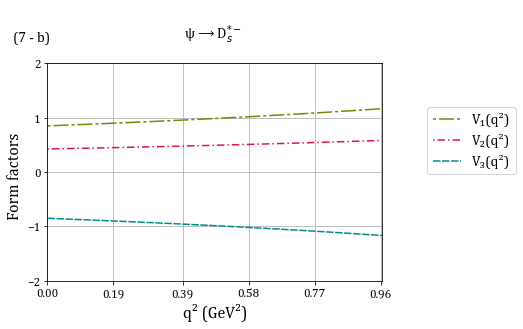}
\includegraphics[width=8.7cm,height=6cm]{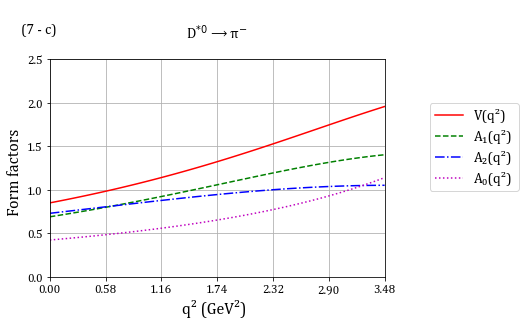}
\includegraphics[width=8.7cm,height=6cm]{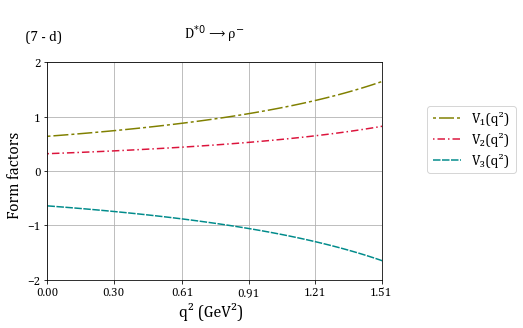}
\includegraphics[width=8.7cm,height=6cm]{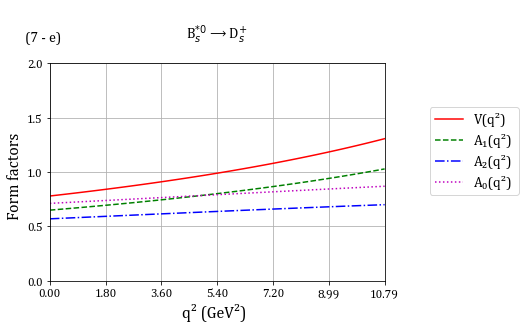}
\includegraphics[width=8.7cm,height=6cm]{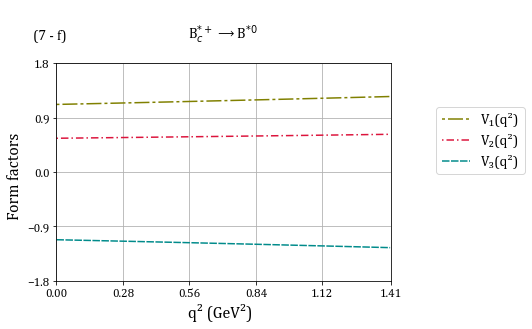}
\caption{The \(q^2\) dependence of the transition form factors for selected decays: (a) \(\psi \to D_s^{-}\), (b) \(\psi \to D_s^{*-}\), (c) \(D^{*0} \to \pi^{-}\), (d) \(D^{*0} \to \rho^{-}\), (e) \(B_s^{*0}\to D_s^{+}\), and (f) \(B_c^{*+}\to B^{*0}\). For \(V \to P\) transitions, the form factors \(V\), \(A_1\), \(A_2\), and \(A_0\) are shown with solid, dashed, dash-dotted, and dotted  lines, respectively. For \(V \to V\) transitions, the form factors \(V_1\), \(V_2\), and \(V_3\) are represented by the custom patterns dot-dot-longdash, dot-space-dash, and dense dash, respectively. }\label{C7}
\end{figure}

Having compared the form factors at zero momentum transfer, we proceed to investigate their behavior across the physical \(q^2\) range. We present results for a representative subset of decays in Figure \ref{C7}, with panels (a) to (f) showing the \(q^2\) dependence for \(\psi \to D_s^{-}\), \(\psi \to D_s^{*-}\), \(D^{*0} \to \pi^{-}\), \(D^{*0} \to \rho^{-}\), \(B_s^{*0}\to D_s^{+}\), and \(B_c^{*+}\to B^{*0}\), respectively. The line styles in these figures distinguish the different form factors: for \(V \to P\) transitions, \(V\) uses a solid line, \(A_{1}\) a dashed line, \(A_{2}\) a dash-dotted line, and \(A_{0}\) a dotted line. For \(V \to V\) transitions, \(V_{1}\) is shown with a dot-dot-longdash pattern, \(V_{2}\) with a dot-space-dash pattern, and \(V_{3}\) with a dense dash pattern.
A distinct advantage of the Hard-Wall AdS/QCD model is its ability to evaluate transition form factors across the full physical $q^2$ range. By solving 5D bulk-to-boundary equations of motion that are continuous in $q^2$, the formalism avoids the kinematic extrapolations and truncated series expansions required by other methods. For instance, LQCD is confined to the small-recoil (large-$q^2$) regime due to signal-to-noise degradation at large spatial momenta, while LCSR is restricted to the large-recoil (small-$q^2$) regime by the convergence limits of the light-cone Operator Product Expansion.
\subsection{Analysis of the branching ratios }
To evaluate the branching ratio values for the $V \to V (P)\,\ell\,\nu_{\ell}$ decays, the decay amplitude in Eq. (\ref{eq.efh}),
and definitions for the form factors given in Eqs. (\ref{eq.deffform}, and \ref{eq.deffform2}) are required. 
For $V\to V\,\ell\,\nu_{\ell}$ decay, the differential decay rate can be written as:
\begin{eqnarray}\label{dewithvv}
\frac{d\Gamma}{dq^2} = \frac{G_F^2 |V_{QQ'}|^2 \sqrt{\lambda(V_i\to V_f)}}{96\pi^3 m_{V_i}^2}\, \frac{q^2}{3}\,   \gamma^2
\times \left[ \frac{3m_\ell^2}{2q^2} \mathcal{V}_{1} + \mathcal{V}_{2} \left(1 + \frac{m_\ell^2}{2q^2} \right) \right]
\end{eqnarray}
where
\begin{eqnarray}
\mathcal{V}_{1}&=&2\left( \frac{m_{V_i}^2 - m_{V_f}^2}{\sqrt{q^2}} V_1 - \sqrt{q^2} V_2 \right)^2 + \mathcal{V}_{3}^2, \label{dewithvv1}\\
\mathcal{V}_{2}&=&2\,\lambda(V_i\to V_f) V_3^2 + 2\left( \frac{2m_{V_i}\sqrt{\lambda(V_i\to V_f)}}{\sqrt{q^2}} V_1 \right)^2 + \mathcal{V}_{3}^2, \label{dewithvv2}\\
\mathcal{V}_{3} &=&\frac{\sqrt{\lambda(V_i\to V_f)}}{\sqrt{q^2} m_{V_f}} \left[ (m_{V_i}^2 + m_{V_f}^2 - q^2) V_1 - \frac{m_{V_i}^2 - m_{V_f}^2 - q^2}{2} V_3 \right], \label{dewithvv3}\\
\lambda(V_i\to V_f)&=&m_{V_i}^{4}+m_{V_f}^{4}+q^4-2\,m_{V_i}^{2}\,m_{V_f}^{2}-2\,m_{V_i}^{2}\,q^2-2\,m_{V_f}^{2}\,q^2,\\ \label{dewithvv4}
\gamma&=&\left(1 - \frac{m_\ell^2}{q^2}\right),
\end{eqnarray}

where $V_i$ and $V_f$ denote the initial and final vector mesons and $V_1$, $V_2$, $V_3$ being the relevant form factors.
The differential decay width for the process \( V \to P\,\ell\,\nu_{\ell} \), expressed in terms of the form factors \( V \), \( A_0 \), \( A_1 \), and \( A_2 \), is given by:
\begin{eqnarray}\label{dewithvp}
\frac{d\Gamma}{dq^2}(V \to P\,\ell\,\nu_{\ell})&=&\frac{G_{F}^{2}\,|V_{QQ'}|^{2}\sqrt{\lambda(V \to P)}}
{384\,m_{V}^{3}\pi^{3}}\frac{\gamma^2}{q^2}\,\Bigg(3\,m_{\ell}^2\,\mathcal{P}_{1}^{2}+\frac{m_{\ell}^2+2\,q^2}{4\,m_P^2}\,\mathcal{P}_{2}^{2}+
(m_{\ell}^2+2\,q^2)\,\mathcal{P}_{3}^{2}\Bigg),
\end{eqnarray}
with
\begin{eqnarray}\label{dewithv}
\mathcal{P}_{1}&=&\sqrt{\lambda(V \to P)}\,A_{0}(q^2),\label{dewithvp11}\\
\mathcal{P}_{2}&=&\Bigg[(m_{V}^{2}-m_{P}^2-q^2)\Bigg((m_{V}+m_{P})\,A_{1}(q^2)+ m_{V}\,m_{P}\frac{2\,V(q^2)}{(m_{V}+m_{P})}\Bigg)-\frac{\lambda\,A_{2}(q^2)}{(m_{V}+m_{P})}\Bigg],\label{dewithvp2}\\
\mathcal{P}_{3}&=&\sqrt{q^2\,\lambda(V \to P)}\Bigg[(m_{V}+m_{P})A_{1}(q^2)+ m_{V}\,m_{P}\frac{2\,V(q^2)}{(m_{V}+m_{P})}\Bigg],\label{dewithvp1}\\
\lambda(V\to P)&=&m_{V}^{4}+m_{P}^{4}+q^4-2\,m_{V}^{2}\,m_{P}^{2}-2\,m_{V}^{2}\,q^2-2\,m_{P}^{2}\,q^2. \label{dewithvp3}
\end{eqnarray}
The numerical inputs used in this analysis are specified as follows. For the CKM matrix elements, we use the values \cite{pdg}:
\begin{eqnarray} \label{ckme}
|V_{cd}| = 0.22, \quad |V_{cs}| = 0.99, \quad |V_{bu}| = 42.2 \times 10^{-3}, \quad |V_{bc}| = 3.94 \times 10^{-3}.
\end{eqnarray}
The total decay widths of the initial mesons are taken from experimental measurements or theoretical predictions. The only meson with available experimental data is the \(\psi\), for which we adopt the experimental value \cite{pdg} 
\begin{eqnarray} \label{wpsi}
\Gamma_{\text{tot}}^{\psi}=(92.6\pm 1.2)~\rm{keV}
\end{eqnarray}
For the vector \(B\) mesons, the total decay widths are approximated by the respective radiative decay widths \(B^* \to B\gamma\) as predicted within 
the SM+CLFQM \cite{CLFQMVV}:
\begin{eqnarray}
\Gamma_{\text{tot}}^{B^{*+}} &\simeq& \Gamma(B^{*+} \to B^{+} \gamma) = (349 \pm 18)\,\text{eV}, \label{widthB1}\\
\Gamma_{\text{tot}}^{B^{*0}}&\simeq& \Gamma(B^{*0} \to B^{0} \gamma) = (116 \pm 6) \, \text{eV}, \label{widthB1}\\
\Gamma_{\text{tot}}^{B_{s}^{*0}} &\simeq& \Gamma(B_{s}^{*0} \to B_{s}^{0} \gamma) = (84^{+11}_{-9}) \, \text{eV},\label{widthB1}\\
\Gamma_{\text{tot}}^{B_{c}^{*+}} &\simeq& \Gamma(B_{c}^{*+} \to B_{c}^{0} \gamma) = (49^{+28}_{-21}) \,\text{eV} \label{widthB1}.
\end{eqnarray}

We note that the radiative decay widths $\Gamma(B^{*} \to B\gamma)$ are directly related to the magnetic coupling $g_{B^{*}B\gamma}$.
While we adopt the predictions from \cite{CLFQMVV} as the central values for our numerical analysis in this study, it is worth emphasizing that this coupling can be rigorously computed from first principles in QCD.  For instance, a refined determination of $g_{B^{*}B\gamma}$ has been recently achieved using the  LCSR incorporating next-to-leading order and higher-twist corrections \cite{Li2020}, providing a highly precise theoretical baseline for these radiative transition widths.

\begin{figure}[th]
\includegraphics[width=8cm,height=7cm]{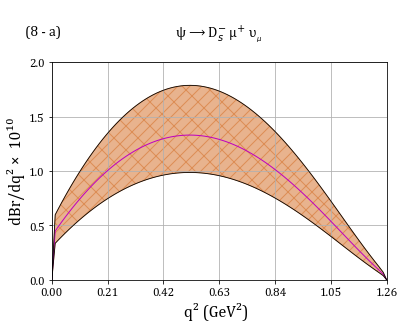}
\includegraphics[width=8cm,height=7cm]{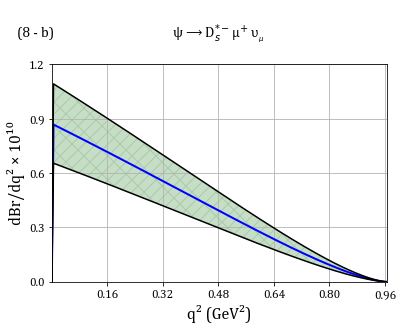}
\includegraphics[width=8cm,height=7cm]{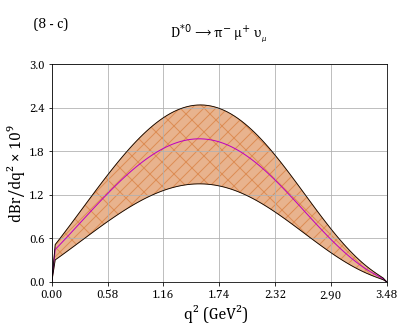}
\includegraphics[width=8cm,height=7cm]{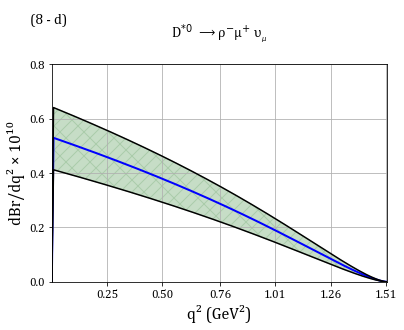}
\includegraphics[width=8cm,height=7cm]{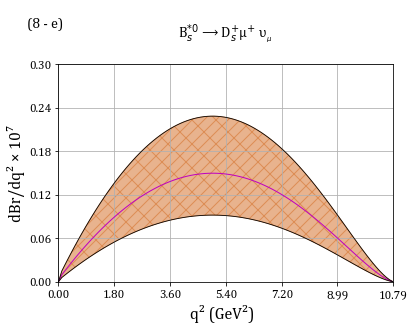}
\includegraphics[width=8cm,height=7cm]{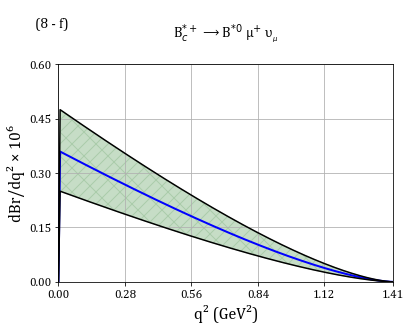}
\caption{ Differential branching ratios \(\text{dBr}/dq^2\) as a function of the dilepton invariant mass squared \(q^2\) for the decays shown in Figure \ref{C7}, plotted for the \(\mu\) channel.  The theoretical predictions are shown as solid lines, with the shaded regions indicating the combined uncertainty.}\label{C8}
\end{figure}

\begin{figure}[th]
\includegraphics[width=8cm,height=7cm]{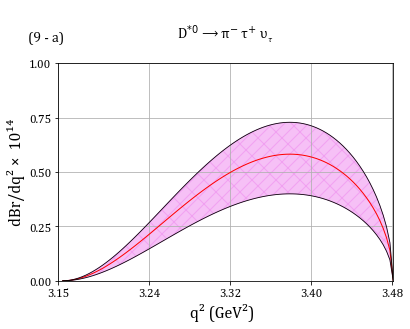}
\includegraphics[width=8cm,height=7cm]{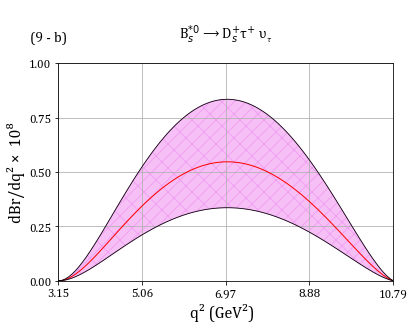}
\caption{The \(q^2\) dependence of the differential branching fraction for the decays \(D^{*0} \to \pi^{-} \tau^{+} \nu_{\tau}\) (left) and \(B_s^{*0} \to D_s^{+} \tau^{+} \nu_{\tau}\) (right). The shaded bands represent the theoretical uncertainties. }\label{C9}
\end{figure}

For the \(D^{*0}\) meson, we use the result from the NF approach \cite{Nfact}:
\begin{eqnarray} \label{wdstar}
\Gamma_{\text{tot}}^{D^{*0}}=(55.9^{+5.9}_{-5.4}) \, \text{keV}
\end{eqnarray}
For the \(D_s^{*+}\) meson, we employ the LQCD prediction \cite{LQCDW}:
\begin{eqnarray} \label{wdstar}
\Gamma_{\text{tot}}^{D_s^{*+}}=(70\pm 28) \, \text{eV}
\end{eqnarray}
Figure \ref{C8} displays the \(q^2\) dependence of the branching ratios for a representative subset of decays in the \(\mu\) channel. The form factors for this subset of decays are shown in Figure \ref{C7} as follows: panel (a) for \(\psi \to D_s^{-}\,\mu^{+} \nu_{\mu}\), panel (b) for \(\psi \to D_s^{*-}\,\mu^{+} \nu_{\mu}\), panel (c) for \(D^{*0} \to \pi^{-}\,\mu^{+} \nu_{\mu}\), panel (d) for \(D^{*0} \to \rho^{-}\,\mu^{+} \nu_{\mu}\), panel (e) for \(B_s^{*0}\to D_s^{+}\,\mu^{+} \nu_{\mu}\), and panel (f) for \(B_c^{*+}\to B^{*0}\,\mu^{+} \nu_{\mu}\).
The differential branching fractions \(d\text{Br}/dq^2\) for the decays \(D^{*0} \to \pi^{-}\tau^{+}\nu_{\tau}\) and \(B_s^{*0} \to D_s^{+}\tau^{+}\nu_{\tau}\) are presented as functions of \(q^2\) in Figs.in Figs. (\ref{C9}- a) and (\ref{C9}- b), respectively. In both Fig. \ref{C8} and Fig. \ref{C9}, the uncertainty regions are indicated by shaded areas.

The branching ratios for the semileptonic decays are obtained by integrating Equations (\ref{dewithvv})
 and (\ref{dewithvp}) over the full physical range of \( q^2 \).  Our estimates for the decay $\psi \to V(P) \ell^+ \nu$ are presented in Table~\ref{TBR1}, along with predictions from the 3PSR~\cite{Wang3PSR}, the CCQM~\cite{Ivanov2015CCQM}, the CLFQM~\cite{CLFQM20242}, and LQCD~\cite{LQCD2008}. Only the $\ell = e, \mu$ channels are considered for the decays in this table.   As can be seen, our results are in good agreement with those of the 3PSR ~\cite{Wang3PSR} method. Additionally, for the 
 $\psi \to D_{s}^{-} \ell^+ \nu$ decay, we also find good agreement with the LQCD~\cite{LQCD2008} results.
 
The branching ratio values  for the decays $V \to P \ell^+ \nu$ are listed in Table~\ref{TBR2}. For the allowed channels, our predictions are included for $\ell = e, \mu, \tau$, along with the predictions from the CLFQM~\cite{CLFQM2024} and the SM~\cite{BtoD2016}.
As can be seen from Table~\ref{TBR2}, when the uncertainty regions of the branching ratios are taken into account, good agreement is found with the results of the CLFQM~\cite{CLFQM2024} and the SM~\cite{BtoD2016} for all of the decays.

Additionally, our results for the decays $B^{*} \to V \ell^{+} \nu_{\ell} (\ell^{-} \bar{\nu}_{\ell})$ with $\ell = \mu, \tau$ are presented in Table~\ref{TBR3}. For better comparison, this table also includes predictions from the CLFQM~\cite{CLFQMVV}, the BS model~\cite{BSS}, and HQS~\cite{HQS}. For most decays listed in Table~\ref{TBR3}, our results agree well with the CLFQM~\cite{CLFQMVV} and the BS model~\cite{BSS} when the respective uncertainty regions are considered.\\
For some of the decays considered, there are no theoretical predictions for the branching ratios. Our estimations yield the following values:  
$\rm{Br} (D^{*0} \to \rho^{-}\,\mu^{+} \nu_{\mu})=(1.67^{+0.36}_{-0.38})\times 10^{-11}$ and  
$\rm{Br} (D^{*0} \to K^{*-}\,\mu^{+} \nu_{\mu})=(2.23^{+0.48}_{-0.51})\times 10^{-10}$.  For $\bar{B}^{*0} \to \rho^{+}\,\mu^{-} \bar{\nu}_{\mu}$, the branching ratio is obtained as $(1.04^{+0.15}_{-0.16})\times 10^{-9}$, while for the $\tau$ channel we find  
$\rm{Br} (\bar{B}^{*0} \to \rho^{+}\,\tau^{-} \bar{\nu}_{\tau})=(2.26^{+0.32}_{-0.35})\times 10^{-10}$.  For $\bar{B}_s^{*0} \to K^{*+}\,\mu^{-} \bar{\nu}_{\mu}$, the branching ratio is $(0.68^{+0.11}_{-0.12})\times 10^{-10}$, while for the $\tau$ channel our analysis gives  
$\rm{Br} (\bar{B}_s^{*0} \to K^{*+}\,\tau^{-} \bar{\nu}_{\tau})=(0.14^{+0.02}_{-0.03})\times 10^{-10}$.

\begin{table}[!th]
\centering
\caption{Estimates for the decay $\psi \to V(P) \ell^+ \nu_{\ell}$ with $\ell =e, \mu$ obtained in this work, together with predictions from the 3PSR~\cite{Wang3PSR}, the CCQM~\cite{Ivanov2015CCQM}, the CLFQM~\cite{CLFQM20242}, and LQCD~\cite{LQCD2008}. }\label{TBR1}
\begin{ruledtabular}
\begin{tabular}{|c|c|c|c|c|c|c|}
Decay Mode  &Unit&3PSR\cite{Wang3PSR} & CCQM \cite{Ivanov2015CCQM}& CLFQM \cite{CLFQM20242} &LQCD \cite{LQCD2008}&This work  \\
\hline
$ \psi \to D^{-}\,\mu^{+}\,\nu_{\mu}$&$10^{-12}$&$7.1^{+4.2}_{-2.2}$&$16.6$&$57.8$&$11.8$&$5.77^{+1.29}_{-1.45}$\\
\hline
$ \psi \to D^{-}\,e^{+}\,\nu_{e}$&$10^{-12}$&$7.3^{+4.3}_{-2.2}$&$17.1$&$61$&$12.1$&$5.82^{+1.29}_{-1.44}$\\
\hline
$ \psi \to D_s^{-}\,\mu^{+}\,\nu_{\mu}$&$10^{-10}$&$1.7^{+0.7}_{-0.5}$&$3.2$&$9.59$&$1.84$&$1.14^{+0.42}_{-0.35}$\\
\hline
$ \psi \to D_s^{-}\,e^{+}\,\nu_{e}$&$10^{-10}$&$1.8^{+0.7}_{-0.5}$&$3.3$&$10.21$&$1.90$&$1.15^{+0.42}_{-0.34}$\\
\hline
$ \psi \to D^{*-}\,\mu^{+}\,\nu_{\mu}$&$10^{-11}$&$3.6^{+1.6}_{-1.1}$&$2.9$&$-$&$-$&$2.81^{+0.51}_{-0.44}$\\ 
\hline
$ \psi \to D^{*-}\,e^{+}\,\nu_{e}$&$10^{-11}$&$3.7^{+1.6}_{-1.1}$&$3.0$&$-$&$-$&$2.87^{+0.52}_{-0.45}$\\
\hline
$\psi \to D_s^{*-}\,\mu^{+}\,\nu_{\mu}$&$10^{-10}$&$5.4^{+1.6}_{-1.5}$&$4.8$&$-$&$-$&$5.19^{+0.83}_{-0.62}$\\
\hline
$ \psi \to D_s^{*-}\,e^{+}\,\nu_{e}$&$10^{-10}$&$5.6^{+1.6}_{-1.6}$&$5.0$&$-$&$-$&$5.21^{+0.84}_{-0.62}$\\
\end{tabular}
\end{ruledtabular}
\end{table}
\begin{table}[!th]
\centering
\caption{Our predictions for the allowed decays $V \to P \ell^{+} \nu_{\ell} (\ell^{-} \bar{\nu}_{\ell})$ with $\ell = e, \mu, \tau$, together with predictions from the CLFQM~\cite{CLFQM2024} and the SM~\cite{BtoD2016}.}\label{TBR2}
\begin{ruledtabular}
\begin{tabular}{|c|c|c|c|c|c|c|c|}
Decay Mode  & Unit &CLFQM \cite{CLFQM2024} &This work& Decay Mode &Unit&CLFQM \cite{CLFQM2024} &This work  \\
\hline
\hline
$ D^{*0} \to \pi^{-}\,\mu^{+}\,\nu_{\mu}$&$10^{-9}$&$2.64^{+0.47}_{-0.29}$&$2.92^{+0.51}_{-0.37}$&$ D^{*0} \to \pi^{-}\,e^{+}\,\nu_{e}$&$10^{-9}$&$2.79^{+0.51}_{-0.30}$&$2.96^{+0.47}_{-0.30}$\\
\hline
$ D^{*0} \to \pi^{-}\,\tau^{+}\,\nu_{\tau}$&$10^{-14}$&$4.96^{+0.67}_{-0.55}$&$5.02^{+0.71}_{-0.47}$&$ D^{*0} \to K^{-}\,\mu^{+}\,\nu_{\mu}$&$10^{-9}$&$7.93^{+0.91}_{-1.03}$&$7.82^{+0.84}_{-0.72}$\\
\hline
$ D^{*0} \to K^{-}\,e^{+}\,\nu_{e}$&$10^{-9}$&$7.93^{+0.91}_{-1.06}$&$7.83^{+0.67}_{-0.56}$&$ D_s^{*0} \to K^{0}\,\mu^{+}\,\nu_{\mu}$&$10^{-7}$&$1.87^{+0.26}_{-0.35}$&$1.56^{+0.43}_{-0.32}$\\
\hline
\hline
Decay Mode  &Unit&SM \cite{BtoD2016} &This work& Decay Mode &Unit&SM \cite{BtoD2016} &This work  \\
\hline
$ B^{*0} \to D^{+}\,\mu^{-}\,\bar{\nu}_{\mu}$&$10^{-8}$&$7.20^{+3.7}_{-3.04}$&$5.82^{+0.92}_{-0.85}$&$ B^{*0} \to D^{+}\,\tau^{-}\,\bar{\nu}_{\tau}$&$10^{-8}$&$2.14^{+1.00}_{-0.89}$&$1.28^{+0.59}_{-0.48}$\\
\hline
$ B_s^{*0} \to D_s^{+}\,\mu^{-}\,\bar{\nu}_{\mu}$&$10^{-7}$&$1.39^{+0.91}_{-0.69}$&$1.06^{+0.48}_{-0.37}$&$B_s^{*0} \to D_s^{+}\,\tau^{-}\,\bar{\nu}_{\tau}$&$10^{-8}$&$4.08^{+2.66}_{-2.03}$&$3.25^{+0.52}_{-0.46}$\\
\end{tabular}
\end{ruledtabular}
\end{table}

\begin{table}[!th]
\centering
\caption{Our predictions for the branching ratios of the decays $B^{*} \to V \ell^{+} \nu_{\ell} (\ell^{-} \bar{\nu}_{\ell})$ with $\ell = \mu, \tau$ are presented alongside predictions from the CLFQM~\cite{CLFQMVV}, the BS model~\cite{BSS}, and HQS~\cite{HQS} for comparison.}\label{TBR3}
\begin{ruledtabular}
\begin{tabular}{|c|c|c|c|c|c|}
Decay Mode  &Unit& CLFQM \cite{CLFQMVV} & BS \cite{BSS}& HQS \cite{HQS} &This work  \\
\hline
$ B^{*-} \to D^{*0}\,\mu^{-}\,\bar{\nu}_{\mu}$&$10^{-8}$&$8.42^{+0.79}_{-0.90}$&$9.36^{+1.04}_{-0.88}$&$6.41$&$7.27^{+1.02}_{-0.98}$\\
\hline
$ B^{*-} \to D^{*0}\tau^{-}\,\bar{\nu}_{\tau}$&$10^{-8}$&$2.26^{+0.23}_{-0.25}$&$2.03^{+0.23}_{-0.20}$&$1.29$&$1.25^{+0.62}_{-0.43}$\\
\hline
$ B^{*0} \to D^{*+}\,\mu^{-}\,\nu_{\mu}$&$10^{-7}$&$2.51^{+0.24}_{-0.26}$&$-$&$1.92$&$2.07^{+0.31}_{-0.27}$\\
\hline
$ B^{*0} \to D^{*+}\,\tau^{-}\,\nu_{\tau}$&$10^{-8}$&$6.73^{+0.68}_{-0.76}$&$-$&$3.88$&$5.21^{+0.42}_{-0.39}$\\
\hline
$ B^{*0}_{s} \to D^{*+}_{s}\,\mu^{-}\,\bar{\nu}_{\mu}$&$10^{-7}$&$3.46^{+0.63}_{-0.68}$&$5.69^{+0.62}_{-0.53}$&$2.53$&$3.26^{+0.45}_{-0.42}$\\ 
\hline
$ B^{*0}_{s} \to D^{*+}_{s}\,\tau^{-}\,\bar{\nu}_{\tau}$&$10^{-8}$&$9.10^{+1.79}_{-1.92}$&$10.30^{+1.50}_{-1.3}$&$5.05$&$4.05^{+0.52}_{-0.49}$\\
\hline
$B^{*+}_{c} \to \psi\,\mu^{+}\,\nu_{\mu}$&$10^{-7}$&$5.44^{+4.46}_{-2.49}$&$11.3^{+1.10}_{-1.00}$&$2.91$&$4.42^{+0.55}_{-0.48}$\\
\hline
$ B^{*+}_{c} \to \psi\,\tau^{+}\,\nu_{\tau}$&$10^{-7}$&$1.43^{+1.22}_{-0.68}$&$3.13^{+0.32}_{-0.28}$&$0.56$&$1.32^{+0.26}_{-0.18}$\\
\hline
$B^{*+}_{c} \to B^{*0}\,\mu^{+}\,\nu_{\mu}$&$10^{-7}$&$-$&$1.59^{+0.21}_{-0.21}$&$-$&$1.25^{+0.31}_{-0.29}$\\
\hline
$ B^{*-}_{c} \to D^{*0}\,\mu^{-}\,\bar{\nu}_{\mu}$&$10^{-9}$&$-$&$4.61^{+0.96}_{-0.76}$&$-$&$3.22^{+0.51}_{-0.49}$\\
\hline
$ B^{*-}_{c} \to D^{*0}\, \tau^{-}\,\bar{\nu}_{\tau}$&$10^{-9}$&$-$&$3.22^{+0.61}_{-0.51}$&$-$&$2.05^{+0.31}_{-0.29}$\\
\end{tabular}
\end{ruledtabular}
\end{table}

\section{Conclusion}\label{sec.5}
In this work, we have performed a comprehensive analysis of semileptonic decays of heavy vector mesons within the Hard-Wall AdS/QCD framework. Unlike pseudoscalar mesons, which have been extensively studied both theoretically and experimentally, the decays of the heavy vector mesons such as \(B^{*}\), \(D^{*}\), \(D_{s}^{*}\), \(B_{c}^{*}\), and the charmonium state \(\psi\) remain comparatively underexplored. Their weak decays are rare due to the small coupling of the weak interaction, and the transition form factors-which encode the non-perturbative strong interaction dynamics-are more numerous and richer in structure because of the additional spin degrees of freedom. The Hard-Wall AdS/QCD model offers an elegant and analytically tractable approach to computing these form factors, as it maps the strongly coupled four-dimensional QCD onto a weakly coupled five-dimensional gravitational theory in AdS space, with an infrared cutoff \(z_0\) that mimics confinement.

We have investigated four distinct quark-level transition categories: \(c \to d\), \(b \to c\), \(c \to s\), and \(b \to u\). For each category, we have analyzed all relevant decay modes, including both vector-to-vector (\(V \to V\)) and vector-to-pseudoscalar (\(V \to P\)) transitions. The complete list of decays studied is: \( \psi \to D^{*-} ( D^{-} ) \ell^{+} {\nu}_{\ell} \), \( D^{*0} \to \pi^{-} \ell^{+} {\nu}_{\ell} \), \( D^{*+}_{s} \to K^{0} \ell^{+} {\nu}_{\ell} \), \( D^{*0} \to \rho^{-} \ell^{+} {\nu}_{\ell} \), \( B_c^{*+} \to B^{*0} \ell^{+} {\nu}_{\ell} \),  \( B_c^{*+} \to \psi \,\ell^{+} {\nu}_{\ell} \), \( B^{*-} \to D^{*0} \ell^{-} \bar{\nu}_{\ell} \), \( B^{*0} \to D^{*+}(D^{+}) \ell^{-} \bar{\nu}_{\ell} \), \( B^{*0}_{s} \to D^{*+}_{s} (D^{+}_{s}) \ell^{-} \bar{\nu}_{\ell} \), \( \psi \to D^{*-}_{s}(D^{-}_{s}) \ell^{+} {\nu}_{\ell} \), \( D^{*0} \to K^{*-} (K^{-}) \ell^{+} {\nu}_{\ell} \), \( B_c^{*-} \to D^{*0} \ell^{-} \bar{\nu}_{\ell} \), \( \bar{B_{s}}^{*0} \to K^{*+} \ell^{-} \bar{\nu}_{\ell} \), and \( \bar{B}^{*0} \to \rho^{+} \ell^{-} \bar{\nu}_{\ell} \).
For each process, we derived the transition form factors as functions of the momentum transfer \(q^{2}\) using the holographic prescription. The wave functions and decay constants of the initial and final mesons were obtained within the same AdS/QCD setup, ensuring theoretical consistency. From the form factors, we evaluated both differential decay rates \(d\Gamma/dq^{2}\) and the integrated branching ratios.

Our numerical results reveal several notable features. First, the form factors exhibit a smooth dependence on \(q^{2}\) across the entire kinematic range, with a pronounced enhancement at low \(q^{2}\) for decays involving light final mesons (e.g., \(D^{*0} \to \pi^{-}\ell\bar{\nu}\)) and a flatter behavior for heavy-to-heavy transitions (e.g., \(B^{*0} \to D^{*+}\ell\bar{\nu}\)). Second, the branching ratios span a wide range, from approximately \(10^{-14}\) up to \(10^{-7}\). 
To assess the reliability of our predictions, we performed a detailed comparison with results obtained from other non-perturbative methods.  Overall, our AdS/QCD results are in good agreement with the majority of these approaches, particularly in the low-\(q^{2}\) region. Discrepancies, where they occur, are typically within (20-30)\% and can be attributed to differences in the treatment of confinement and the extrapolation to high \(q^{2}\).

\end{document}